\documentclass[aip,jcp,longbibliography]{revtex4-1}
\usepackage{epsfig}
\usepackage{textcomp}
\usepackage{graphicx}
\usepackage{amsmath}
\usepackage{amssymb}
\usepackage{url}
\usepackage[normalem]{ulem}
\usepackage{booktabs}

\usepackage{color}
\usepackage{soul}
\definecolor{Myblue}{rgb}{0.3,0.3,1.0}
\usepackage{xr}
\usepackage{bm}

 \usepackage{rotating}

\newcommand*{\citen}[1]{%
  \begingroup
    \romannumeral-`\x % remove space at the beginning of \setcitestyle
    \setcitestyle{numbers}%
    \cite{#1}%
  \endgroup   
}

\begin{document}

\title{Machine Learning for Observables: Reactant to Product State
  Distributions for Atom-Diatom Collisions}

\author{Julian Arnold, Debasish Koner, Silvan K\"aser}
\affiliation{Department of Chemistry, University of Basel,
  Klingelbergstrasse 80, CH-4056 Basel, Switzerland}

\author{Narendra Singh} \affiliation{Department of Mechanical
  Engineering, Stanford University, CA 94305 USA}

\author{Raymond J. Bemish} \affiliation{Air Force Research Laboratory,
  Space Vehicles Directorate, Kirtland AFB, New Mexico 87117, USA}

\author{Markus Meuwly}\email{m.meuwly@unibas.ch}
\affiliation{Department of Chemistry, University of Basel,
  Klingelbergstrasse 80, CH-4056 Basel, Switzerland}

\date{\today}

\begin{abstract}
Machine learning-based models to predict product state distributions
from a distribution of reactant conditions for atom-diatom collisions
are presented and quantitatively tested. The models are based on
function-, kernel- and grid-based representations of the reactant and
product state distributions. While all three methods predict final
state distributions from explicit quasi-classical trajectory
simulations with $R^2 > 0.998$, the grid-based approach performs
best. Although a function-based approach is found to be more than two
times better in computational performance, the kernel- and grid-based
approaches are preferred in terms of prediction accuracy,
practicability and generality. The function-based approach also
suffers from lacking a general set of model functions. Applications of
the grid-based approach to non-equilibrium, multi-temperature initial
state distributions are presented, a situation common to energy
distributions in hypersonic flows. The role of such models in Direct
Simulation Monte Carlo and computational fluid dynamics simulations is
also discussed.
\end{abstract}

\maketitle

\section{Introduction}
The realistic description of chemical and reactive systems with a
large number of available states, such as explosions, hypersonic gas
flow around space vehicles upon re-entry into the atmosphere, or
meteorites penetrating deep into the lower parts of Earth's or a
planet's atmosphere, requires an understanding of the relevant
processes at a molecular
level.\cite{park:1993,cummings:2003,MM.hypersonics:2020} Correctly
describing the population of the available state space under such
non-equilibrium conditions (e.g. high temperatures in atmospheric
re-entry with $T > 10000$ K) from ground-based experiments is
extremely challenging. Such high gas temperatures make the gathering
of experimental data exceedingly difficult but is essential for
simulations of hypersonic flight.\cite{boyd:2015} On the other hand, a
comprehensive modeling of gas phase chemical reactions through
explicit molecular-level simulations remains computationally
challenging due to the large number of accessible states and
transitions between them.\cite{grover:2019,MM.nncs:2019} There are
also other, similar situations in physical chemistry such as the
spectroscopy in hot environments (e.g. on the sun) for which small
polyatomic molecules can populate a large number of rovibrational
states\cite{HITRAN} between which transitions can take
place. Exhaustively probing and enumerating all allowed transitions or
creating high-dimensional analytical representations for them is
usually not possible. Nevertheless, it is essential to have complete
line lists available because, if specific states that are involved in
important transitions are omitted, modeling of the spectroscopic bands
becomes difficult or even impossible.\cite{tennyson:2014} This points
towards an important requirement for such models, namely that they
contain the majority of the important information while remaining
sufficiently fast to evaluate.\\
 
\noindent
In such situations, machine learning approaches can provide an
alternative to address the problem of characterizing product
distributions from given reactant state distributions. In previous
work\cite{MM.nncs:2019}, a model for state-to-state (STS) cross
sections of an atom-diatom collision system using a neural network
(NN) has been proposed. Motivated by the success of such an approach,
the present work attempts to develop a NN-based
distribution-to-distribution (DTD) model for the relative
translational energy $E_{\rm trans}$, the vibrational $v$ and
rotational $j$ states of a reactive atom-diatom collision system. In
other words, given the reactant state distributions
($P(E_{\text{trans}}), P(v), P(j)$) such a model predicts the three
corresponding product state distributions ($P(E_{\text{trans}}'),
P(v'), P(j')$). Here, $P(v)$ and $P(j)$ are marginal distributions,
i.e. $P(v)=\sum_{j} P(v,j)$ and $P(j)=\sum_{v} P(v,j)$, where $(v,j)$
labels the rovibrational state of the diatom.\cite{schwartz:2018}
Hence, instead of considering all possible combinations of
($E_{\text{trans}}, v, j$) on the reactant (input) and product
(output) side explicitly, one is rather interested in a description of
these microscopic quantities by means of their underlying probability
distributions.\\

\noindent
While the state-to-state specificity is lost, such a probabilistic
approach considerably reduces the computational complexity of the
problem at hand. While a STS-approach is still feasible for an
atom-diatom collision system with $\sim 10^{7}$ STS cross
sections\cite{MM.nncs:2019}, it becomes
intractable\cite{schwartzentruber:2018} even for a diatom-diatom type
collision system due to the dramatic increase in the number of STS
cross sections to $\sim 10^{15}$. Moreover, such DTD models still
allow for the prediction of quantities relevant to hypersonic flow,
such as the reaction rates or the average vibrational and rotational
energies.\cite{singh2:2019}\\

\noindent
Here, the N + O$_{2}\rightarrow$ NO + O reaction is considered, which
is relevant in the hypersonic flight regime and for which accurate,
fully dimensional potential energy surfaces (PESs) are
available.\cite{MM.no2:2020} The necessary reference data to train the
NN-based models was obtained by running explicit quasi-classical
trajectories (QCT) for reactive N + O$_2$ collisions. In particular,
from a diverse set of equilibrium reactant state distributions
($P(E_{\text{trans}}), P(v), P(j)$) for N + O$_{2}$, the corresponding
product distributions for NO + O are obtained by means of QCT
simulations. In this work, three different approaches for learning and
characterizing these distributions are pursued, including function-,
kernel-, and grid-based models (F-, K-, and G-based models in the
following). The microscopic description provided by such DTD models
can, e.g., be used as an input or to develop models for more
coarse-grained approaches, including Direct Simulation Monte
Carlo\cite{dsmc} (DSMC) or computational fluid dynamics (CFD)
simulations. Furthermore, the core findings of this work also carry
over to applications in other areas where a DTD model is of interest,
such as in demographics\cite{flaxman:2016} or
economics\cite{perotti:1996}.\\

\noindent
This work is structured as follows. First, the methods including three
different approaches to construct NN-based DTD models are
described. Then, the performance of the models is assessed for various
data sets and improvements in particular related to the input features
are explored. Finally, implications for modeling hypersonic gas flow,
based on DSMC and CFD simulations, are discussed and conclusions are
drawn.\\

\section{Methods}
\subsection{Quasi-Classical Trajectory Calculations}
\label{QCT}
\noindent
Explicit QCT simulations for the N + O$_{2}$ collision system were
carried out on the $^4$A$'$ PES of NO$_2$ following previous
work.\cite{tru79,hen11,kon16:4731,MM.cno:2018,MM.no2:2020}
Specifically, the reactive channel for the N + O$_2$ $\rightarrow$ NO
+ O collision was considered. The $^4$A$'$ PES is chosen here, because
this state contributes most to the equilibrium rate.\cite{MM.no2:2020}
Briefly, Hamilton's equations of motion are solved in reactant Jacobi
coordinates using a fourth-order Runge-Kutta method with a time step
of $\Delta t = 0.05$~fs, which guarantees conservation of the total
energy and angular momentum. The initial conditions for each
trajectory were randomly chosen using standard Monte Carlo sampling
methods.\cite{tru79,hen11} The initial relative translational energies
$E_{\rm trans}$ were sampled from Maxwell-Boltzmann distributions
($E_{\rm trans,min} = 0.0$ eV; $E_{\rm trans,max} = 19.8$ eV;$\Delta
E_{\rm trans} = 0.1$ eV) and reactant vibrational $v$ and rotational
$j$ states were sampled from Boltzmann distributions ($v_{\rm min}=0,
v_{\rm max}=38, \Delta v = 1$; $j_{\rm min}=0, j_{\rm max}=242, \Delta
j = 1$), characterized by $T_{\text{trans}}$, $T_{\text{vib}}$ and
$T_{\text{rot}}$, respectively.\cite{tru79,bender:2015} The impact
parameter $b$ was sampled from 0 to $b_{\rm max} = 10$ \AA\/ using
stratified sampling\cite{tru79,bender:2015} with 6 equidistant strata.
The rovibrational reactant (O$_{2}$; $(v,j)$) and product diatom (NO;
$(v',j')$) states are calculated following semiclassical theory of
bound states.\cite{kar65:3259} The states of the product diatom are
assigned using histogram binning.\\

\noindent
First, models were constructed for the case
$T_{\text{rovib}}=T_{\text{rot}}=T_{\text{vib}}$, for which QCT
simulations were performed at $T_{\text{trans}}$ and
$T_{\text{rovib}}$ ranging from 5000 K to 20000 K in increments of 250
K. This yielded 3698 sets of reactant state distributions and
corresponding product state distributions which will be referred to as
``Set1''. Next, for the more general case $T_{\text{rot}} \neq
T_{\text{vib}}$, further QCT simulations were performed for
$T_{\text{trans}} = 5000, 10000, 15000, 20000$ K with $T_{\text{vib}}$
and $T_{\text{rot}}$ each ranging from 5000 K to 20000 K in increments
of 1000 K. This gives an additional 960 data sets and a total of 4658
data sets that include both cases, $T_{\text{rot}} = T_{\text{vib}}$
and $T_{\text{rot}} \neq T_{\text{vib}}$, collectively referred to as
``Set2''.\\

\noindent
The reactant and product state distributions of Set1 and Set2
constitute representative reference data to train and validate
NN-based models. For both sets the temperatures $\bm{T}=
(T_{\text{trans}}$, $T_{\text{vib}}$, $T_{\text{rot}}$) completely
specify the reactant and product state distributions as they are
related through the explicit QCT simulations. Hence, for brevity a
specific set of reactant and product state distributions is referred
to as $\bm{T}$.\\

\subsection{Generating Nonequilibrium Data Sets}
\label{Generating_noneq_Datasets}
In hypersonic applications it is known that quantities such as
$P(E_{\text{trans}})$, $P(v)$, or $P(j)$ are typically nonequilibrium
probability distributions.\cite{boyd:2015} This has also been
confirmed in explicit simulation studies, starting from equilibrium
energy and state distributions as is commonly done in QCT
simulations.\cite{alp17:2392,MM.cno:2018} Therefore, a general DTD model
should be able to correctly predict (nonequilibrium) product state
distributions starting from nonequilibrium reactant state
distributions. With this in mind, and with Set2 at hand, new reactant
and product state distributions were generated by means of a weighted
sum of the existing distributions according to
\begin{equation}
P(i)=\frac{1}{w_{\text{tot}}}\sum_{n=1}^N w_n \cdot P_{n}(i).
\label{eq:multit}
\end{equation}
Here, $i \in [E_{\text{trans}},v,j,E_{\text{trans}}',v',j']$ labels
the degree of freedom, $n$ labels the data set, $N \in [2,3]$ is the
total number of distributions drawn from Set2, the corresponding
distributions $P_{n}(i)$ used for and obtained from QCT simulations
and the random weights $w_{n} \in [1,2]$ determine how much these
contribute to the total sum. The resulting distributions are scaled by
$w_{\text{tot}}=\sum_{n=1}^N w_n$ to conserve probability. Such
distributions then constitute Set3. It is assumed that any
nonequilibrium state distribution can be represented as a
decomposition in terms of a linear combination of equilibrium
distributions given by Eq.~\ref{eq:multit}. For instance, general
nonequilibrium distributions for nitrogen and oxygen relaxation at
high temperature ($> 8000$ K) conditions, obtained via direct
molecular simulations (DMS), which is equivalent to solving the full
master equation, have been successfully modelled as a weighted sum of
two Boltzmann distributions.\cite{singh3:2019} Consequently, DTD
models that are successfully trained and validated on Set3 are also
expected to generalize well to most nonequilibrium situations
encountered in practice.\\

\noindent
In the following, a single data set for Set3 is generated by randomly
specifying the number of distributions $N \in [2,3]$ to be combined
although larger values for $N$ are possible and will be explored
later. The final set of reactant and product state distributions is
characterized by $N$ sets of temperatures $\bm{T}$ and corresponding
normalized weights $\bm{w}=(w_{1}/w_{\rm tot},...,w_{N}/w_{\rm
  tot})$. The product state distributions obtained by this procedure
are akin to explicit QCT simulations using Monte Carlo sampling of the
reactant state distributions by sampling each of the equilibrium
distributions in the corresponding weighted sum. In the following,
three different possibilities for characterizing reactant and product
state distributions are described.\\

\subsection{Function (F)-Based Approach}
\begin{figure}[h!]
\begin{center}
\includegraphics[width=0.99\textwidth]{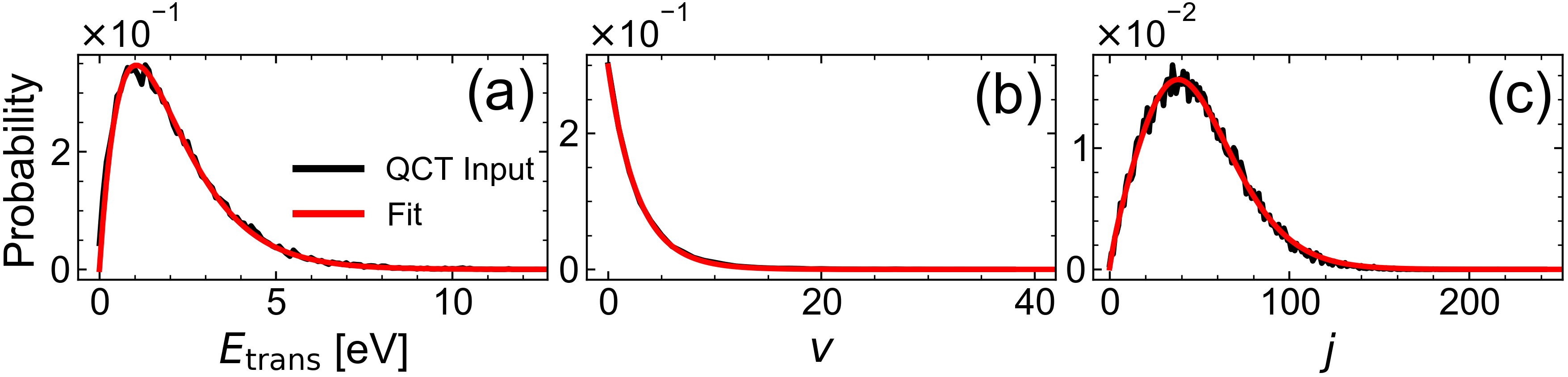}
\caption{Reactant state distributions for the F-based approach: $P(E_{\rm
    trans})$ ($T_{\text{trans}} = 12500$ K, panel (a), $P(v)$ (panel
  (b)) and $P(j)$ (panel (c)) distributions ($T_{\text{rovib}} = 5750$
  K) for explicit QCT simulations (black) and corresponding fits (red)
  obtained using Eqs. \ref{eq:func_E_r} to \ref{eq:func_j_r}.}
\label{fig:fig1}
\end{center}
\end{figure}

In the F-based approach, each set of relative translational energy,
vibrational and rotational state distributions of reactant and product
are fitted to parametrized model functions, see Figure \ref{fig:fig1}
for an example. The corresponding fitting parameters in
Eqs. \ref{eq:func_E_r} to \ref{eq:func_j_p} constitute the input and
output of a NN, respectively (see Section \ref{Neural Network} for
details on the NN). Together with the parametrized model functions
(Eqs. \ref{eq:func_E_r} to \ref{eq:func_j_p}) this serves as a map
between reactant and product state distributions, i.e., a DTD
model. In this work the F-based approach was only applied to Set1.\\

\noindent
The set of model functions used here was
\begin{equation}
\tilde{P}(E_{\text{trans}}) = a_{1} E_{\text{trans}}\cdot
\exp(-E_{\text{trans}}/a_{2}),
\label{eq:func_E_r}
\end{equation}
\begin{equation}
\tilde{P}(v) = b_{1}\exp(-v/b_{2}),
\label{eq:func_v_r}
\end{equation}
\begin{equation}
\tilde{P}(j) = c_{1}\exp(-j/c_{2}) + c_{3}\exp(-(\ln(2c_{4}(j-c_{5})/c_{6}+1)/c_{4})^2),
\label{eq:func_j_r}
\end{equation} 
\begin{equation}
\tilde{P}(E_{\text{trans}}') =
d_{1}\exp(-(\ln(2d_{2}(E_{\text{trans}}'-d_{3})/d_{4}+1)/d_{2})^2),
  \label{eq:func_E_p}
\end{equation}
\begin{equation}
 \tilde{P}(v') = e_{1}v'^4+e_{2}v'^3+e_{3}v'^2+e_{4}v'+e_{5},
  \label{eq:func_v_p}
\end{equation}
\begin{equation}
\tilde{P}(j') = f_{1}\exp(-j'/f_{2}) + f_{3}\exp(-(\ln(2f_{4}(j'-f_{5})/f_{6}+1)/f_{4})^2),
\label{eq:func_j_p}
\end{equation} 
where $\bm{a}=(a_{1},a_{2})$ through $\bm{f}=(f_{1},...,f_{6})$ are
the fitting parameters of the model functions. In total, this results
in 10 and 15 fitting parameters for one set of reactant or product state
distribution, respectively. For the reactant and product rotational
state distributions, $P(j)$ and $P(j')$, the same model function was
used. Such a parametric approach has its foundation in surprisal
analysis\cite{levine:1978} which was recently used in models for
hypersonics.\cite{schwartz:2018,singh1:2019,singh2:2019}\\

\noindent
The reactant state distributions in Set1 are equilibrium distributions
and the model functions (Eqs. \ref{eq:func_E_r} to \ref{eq:func_j_r})
were chosen accordingly. For the product state distributions, which
are typically nonequilibrium distributions, modified parametrizations
were used after inspection of the results from the QCT simulations.
Here, it is worth mentioning that using alternative parametrizations
for model functions are possible, too, which will be briefly explored
later.\\

\subsection{Kernel (K)-Based Approach}
\label{K-based method}
The representer theorem\cite{scholkopf2001generalized} states that,
given $N$ grid points $x_i$, the function $f(x)$ can always be
approximated as a linear combination of suitable functions
\begin{equation}
f(x) \approx \widetilde{f}(x) = \sum_{i = 1}^{N} c_i K(x,x_i)
\label{RKHS}
\end{equation}
where $c_i$ are coefficients and $K(x,x_i)$ is a kernel function. The
reproducing property asserts that $f(x') = \langle f(x),K(x,x')
\rangle$ where $\langle \cdot \rangle$ is the scalar product,
$K(x,x')$ is the kernel\cite{aronszajn:1950} and the coefficients
$c_i$ are determined through matrix inversion. This leads to a
reproducing kernel Hilbert space (RKHS) representation that exactly
reproduces the function at the grid points
$x_i$.\cite{aronszajn:1950,ho96:2584,MM.RKHS:2017} In the present work
the amplitudes of the distributions at the chosen grid points are used
for inter- and extrapolation based on a RKHS-based representation and
the the coefficients $c_{i}$ serve as input and output of the
NN. Hence, given the kernel coefficients for the reactant state
distributions, the NN is trained to predict the coefficients of the
corresponding product state distributions. Together with the
associated grids, one obtains a continuous (K+NN-based) prediction of
the product state distributions, i.e., a DTD model. The K-based
approach was also only applied to Set1.\\

\noindent 
In this work, a Gaussian kernel
\begin{equation}\label{gaussian_ker}
K(x,x')=\exp(-|x-x'|^2/2\sigma^2),
\end{equation}
with hyperparameter $\sigma$ was found to perform well as a
reproducing kernel. Furthermore, a variable amount of regularization
as specified by the regularization rate $\lambda$ in the
Tikhonov-scheme is used. The hyperparameters $\sigma$ and $\lambda$
remain to be optimized systematically. However, the present choices
yielded sufficiently accurate representations. Assigning $\sigma$ to
the average spacing between neighbouring grid points was found to be
advantageous. Alternatively, it is also possible to choose $\sigma$ at
each grid point to be equal to the larger of the two neighbouring grid
spacings. In this work, the first approach is used for $P(v)$ and
$P(v')$, whereas the second approach is applied for all other
distributions. Moreover, such choices for $\sigma$ significantly
reduced the number of grid points required for accurate RKHS
approximations and the resulting kernel coefficients lead to accurate
NN predictions. While the same level of accuracy for the RKHS
approximations can be achieved with larger values for $\sigma$, the
accuracy of the resulting NN predictions was found to
deteriorate. Consequently, only the regularization rate $\lambda$
needed to be tuned and accurate RKHS approximations were obtained
after a few iterations. \\

\begin{figure}[b!]
\begin{center}
\includegraphics[width=0.99\textwidth]{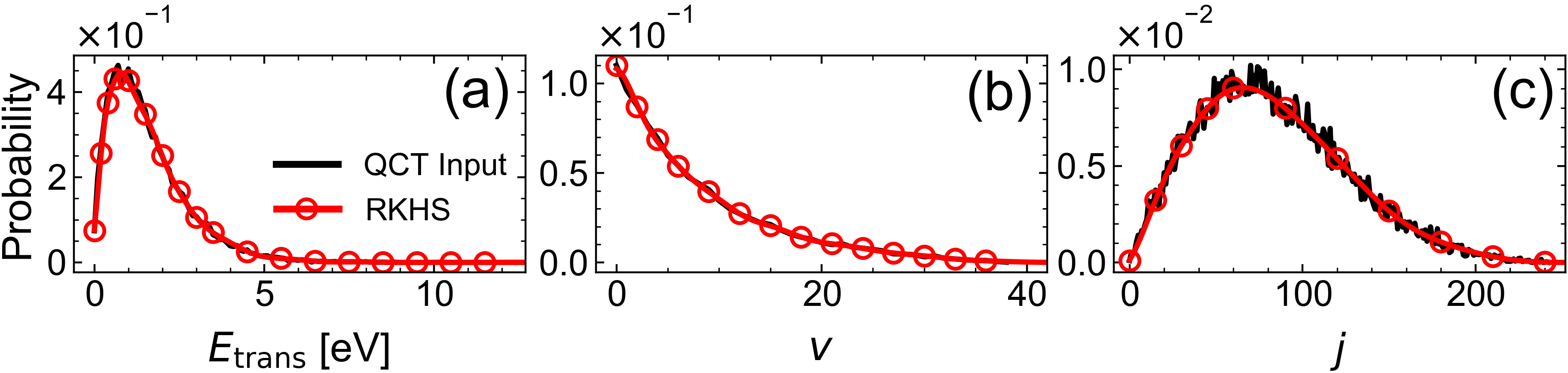}
\caption{Reactant state distributions for K- and G-based approaches: $P(E_{\rm
    trans})$ ($T_{\text{trans}} = 9500$ K, panel (a), $P(v)$ (panel
  (b)) and $P(j)$ (panel (c)) distributions ($T_{\text{rovib}} =
  16000$ K) for explicit QCT simulations (black), their RKHS
  representations (red lines) and the locally averaged values at the
  corresponding grid points (red circles) used for the G-based approach.}
\label{fig:fig2}
\end{center}
\end{figure}

\noindent
The location and number of grid points for the reactant and product
state distributions is largely arbitrary but should be governed by the
overall shape of the distributions, see Figure \ref{fig:fig2} for an
example. The grids used here are reported in Table~S1. The number of grid points for reactant and
product state distributions differs because they are equilibrium and
nonequilibrium distributions, respectively. Also, depending on the
shape of the distributions to be represented, additional points may be
required to avoid unphysical undulations in the RKHS
approximations. For the system considered here, this is mainly
observed for $P(v')$ which requires a denser grid than the
corresponding reactant state distributions $P(v)$.\\

\noindent
Instead of directly evaluating the distributions at the grid points,
local averaging over neighboring data points according to
\begin{equation}
\bar{P}(x_{i}) = \frac{1}{2n+1}\sum\limits_{j=i-n}^{i+n} P(x_{j}),
\label{eq:local_averaging}
\end{equation}
was performed to obtain $\bar{P}(x_{i})$. Here $n = {\rm min}(n_{\rm
  max},n_{\rm nb})$, with $n_{\rm nb}$ the maximum number of
neighbouring data points to the left or the right. If the first and
last data points are chosen as grid points they were assigned the
unaveraged distribution values. The value of $n_{\text{max}}$ can
differ for each of the
($E_{\text{trans}},v,j,E_{\text{trans}}',v',j'$) distributions. For
the K-based approach these values were $n_{\text{max},E_{\rm trans}} =
2$, $n_{\text{max},v} = 1$, $n_{\text{max},j} = 12$ for the reactant
and $n_{\text{max},E_{\rm trans}'} = 3$, $n_{\text{max},v'} = 2$, and
$n_{\text{max},j'} = 13$ for the product state distributions. Local
averaging can be seen as an implicit regularization as it reduces the
noise partially arising due to finite sample statistics in the QCT
simulations.\\

\subsection{Grid (G)-Based Approach}
For the G-based approach, the same grids as in the K-based approach
are considered (see Table~S1). Furthermore,
similar to the K-based approach local averaging is performed with
$n_{\text{max},E_{\rm trans}} = 2$, $n_{\text{max,}v} = 1$,
$n_{\text{max,}j} = 9$ for the reactant and $n_{\text{max},E_{\rm
    trans}'} = 3$, $n_{\text{max,}v'} = 2$, and $n_{\text{max,}j'} =
10$ for the product state distributions. These values were adjusted
such as to obtain accurate discrete representations of the
corresponding distributions. In the G-based approach the locally
averaged values of reactant state distributions at the grid points
(referred to as ``amplitudes'') directly serve as the input of a NN,
see the red circles in Figure \ref{fig:fig2}. The NN then predicts the
product state distributions on the corresponding grids, where the
amplitudes serve as the reference. The resulting discrete product
distributions are finally represented as a continuous RKHS,
establishing a DTD model. Similarly to the K-based approach, a
Gaussian kernel (Eq. \ref{gaussian_ker}) was used for the RKHS of the
product state distributions, as this choice still yielded accurate
approximations. Furthermore, the corresponding hyperparameters
$\sigma$ and $\lambda$ are were chosen as in the K-based approach.\\

\subsection{Neural Network}
\label{Neural Network}
\begin{figure}[h!]
\includegraphics[width=0.35\textwidth]{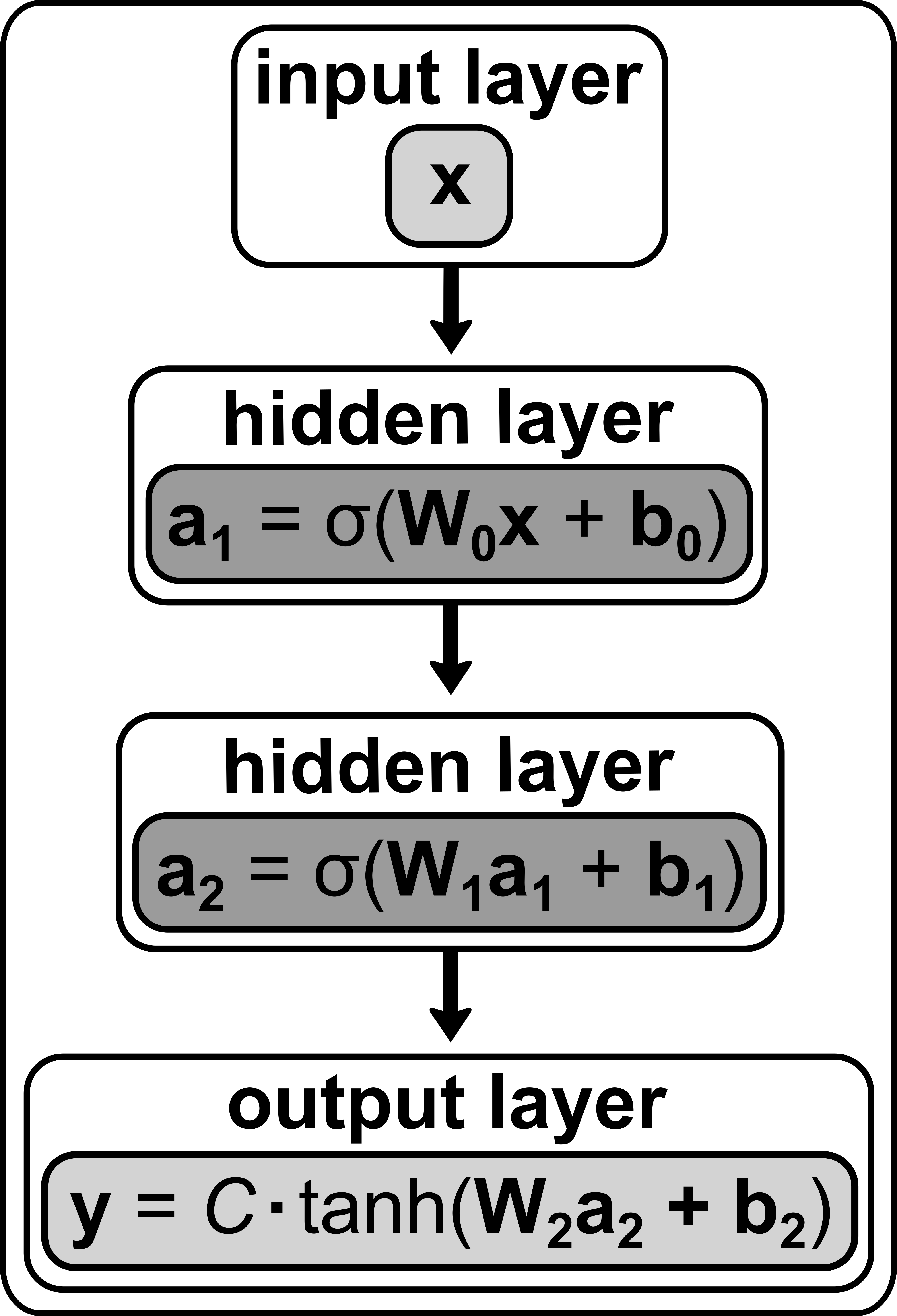}
\caption{Schematic diagram of the NN architecture. The activation
  vector of each hidden layer is \textbf{a$_{i}$}. The input and output
  vectors are $\mathbf{x}$ and $\mathbf{y}$ and the weight matrix and
  bias vector for each layer are $\mathbf{W}_{i}$ and
  $\mathbf{b}_{i}$, respectively. The activation function of the
  hidden layers is $\sigma(\cdot)$ and corresponds to a shifted
  softplus\cite{dugas:2001} function $\sigma(\cdot) =
  \ln(1+e^{(\cdot)})-\ln(2)$. Here, activation functions act
  element-wise on input vectors $(\cdot)$. The constant $\ln(2)$
  centers the mean activation of the hidden nodes at zero, thereby
  decreasing the bias shift effect which speeds up the
  learning.\cite{clevert:2015, lecun:2012} The activation function of
  the output layer is $C \cdot \tanh(\cdot)$\cite{glorot:2011}, where
  $C$ is an overall scaling with $C=8/12/8$ in the F-, K- and G-based
  approaches.}
\label{fig:fig4}
\end{figure}

The NN architecture for training the three models is a multilayer
perceptron with two hidden layers, see Figure \ref{fig:fig4}. The
input and output layers consist of 10/43/43 input and 15/44/44 output
nodes in the F-, K-, and G-based approaches. The input/output are the
fitting parameters (F-based approach), kernel coefficients (K-based
approach), and amplitudes (G-based approach) characterizing reactant
and product state distributions, respectively. When training a NN
using Set1 to Set3 the two hidden layers contain 6, 12, and 9 nodes
each, respectively.\\

\noindent
For training, the input and output of the NN are standardized
according to $x_{i}' = (x_{i} - \bar{x}_{i})/\sigma_{i}$, where
$x_{i}$, $\bar{x}_{i}$ and $\sigma_{i}$ are the $i$-th input/output, and
the mean and standard deviation of their distribution over the entire set of training data. Scaling of the
input, here by means of standardization, is common practice in the
data pre-processing step for machine learning tasks relying on
gradient descent for optimization, as it generally yields a faster
convergence rate.\cite{lecun:2012} The additional standardization of
the output allows one to use a root-mean-square deviation (RMSD) loss
function, as the non-standardized values can differ drastically in
magnitude and spread. Thus, in particular low and high product state
probabilities can be predicted with similar accuracy. It would also be
possible to simply normalize the output but the additional offset
gives the flexibility to use a scaled hyperbolic tangent as
an activation function for the output layer which increases the NN
prediction accuracy compared to other/no activation functions. The
RMSD loss function $\mathcal{L}$ used here is
\begin{equation}
\label{loss_func}
\mathcal{L} = \sqrt{\frac{1}{N}\sum\limits_{i=1}^{N} \left( y_{i} -
  y'_{i}\right)^2},
\end{equation}
with $y_{i}$ and $y'_{i}$ the predicted and reference output values.\\

\noindent
The weights and biases of the NN are initialized according to the
Glorot scheme\cite{glorot2010understanding} and optimized using
Adam\cite{kingma2014adam} with an exponentially decaying learning
rate. The NN is trained using TensorFlow \cite{tf:2016} and the set of
weights and biases resulting in the lowest loss as evaluated on the
validation set are used for predictions. When training a NN using Set1
with $N_{\text{tot}}=3698$ data sets, $N_{\text{train}}=3000$ were
randomly selected for training, $N_{\text{valid}}=600$ for validation
and $N_{\text{test}}=98$ as a test set, whereas for Set2
$N_{\text{tot}}=4658$, $N_{\text{train}}=3600$,
$N_{\text{valid}}=900$, and $N_{\text{test}}=158$.\\

\noindent
To train models on reactant state distributions that can {\it not} be
characterized as a single set of temperatures $\bm{T}$, Set3 was
constructed by means of a weighted sum of the distributions in Set2
(see Section \ref{Generating_noneq_Datasets}). For this, 158 data sets
are randomly selected from $N_{\text{tot}}=4658$ data sets of
Set2. They constitute the subset from which the final test set of Set3
is generated. Here, $N_{\text{test}}=125$, see Section
\ref{Generating_noneq_Datasets}. The remaining 4500 data sets make up
the subset from which the data sets for training and validation are
constructed. In particular, $N_{\text{train}}+N_{\text{valid}}= 5000,
10000,15000, 20000, 25000, 30000$ data sets are generated by means of
the same procedure, making up the final training and validation sets
of Set3 with $N_{\text{train}} = 0.8 \times
(N_{\text{train}}+N_{\text{valid}})$ and $N_{\text{valid}} = 0.2
\times (N_{\text{train}}+N_{\text{valid}})$. All NNs in this work were
trained on a 1.8 GHz Intel Core i7-10510U CPU with training times
shorter than 10 minutes.\\

\section{Results}
The results section first presents DTD models for the F-, K-, and
G-based approaches for $T_{\text{vib}} = T_{\text{rot}}$. This is
followed by discussing the influence of featurization, computational
cost and generalizability of the approaches considered. As the G-based
approach is found to perform best from a number of different
perspectives, models are then trained for $T_{\text{vib}} \neq
T_{\text{rot}}$ and for nonequilibrium  reactant state distributions. Also,
variations of the G-based approach requiring fewer input data are
explored. Then, the findings are discussed in a broader context and
conclusions are drawn.\\

\subsection{Distribution-to-Distribution Models for $T_{\text{vib}} = T_{\text{rot}}$}
\label{Dist_to_Dist}
First, an overall assessment of the three different approaches for
describing the reactant state distributions for Set1
(i.e. $T_{\text{vib}} = T_{\text{rot}}$) is provided. As they are
generated according to the typical sampling procedures followed by QCT
simulations and not from direct function evaluations of equilibrium
distributions they contain noise. This is done because the entire work
is concerned with a situation typically encountered in QCT simulations
of reactive processes.\cite{kar65:3259} Also, it is noted that for the
reactant state distributions one should find - as demonstrated here -
that they are characterized by one parameter only, namely
temperature. This, however, is an open point and not guaranteed for
the product state distributions, which are nonequilibrium
distributions in general.\\

\noindent
For the F- and K-based approaches the reactant state distributions are
first represented either as a parametrized model function or as a
RKHS, respectively, and the agreement is found to be excellent (see
Figures \ref{fig:fig1} and \ref{fig:fig2}). For the G-based approach
this step is not required, as the input are the amplitudes of the
reactant state distributions themselves (see Figure \ref{fig:fig2}).\\

\noindent
Having established that all three (F-, K-, and G-based) approaches are
suitable to describe reactant state distributions, a NN for each
of the three models was trained on Set1 with $N_{\text{train}}=3000$
and $N_{\text{valid}}=600$. The quality of the final model for
predicting product state distributions depends on two aspects: 1. The
ability of the NN to {\it learn} and predict the product state
distributions obtained from the QCT simulations 2. The ability of the
(F-, K- or G-based) approaches to {\it describe} these
distributions.\\

\noindent
{\it 1. Quality of the NN Prediction:} For the first aspect, RMSD and
coefficient of determination ($R^2$) values, referred to as RMSD$_{\rm
  NN}$ and $R^2_{\rm NN}$, are considered as performance measures. For
a single data set from the test set these are calculated by comparing
the normalized reference representations of each of the models and the
corresponding normalized NN predictions on the grid for which QCT data
is available for each of the three product state distributions
separately and averaging over the resulting values. The normalization
factors were calculated by numerical integration of the distributions
obtained from the QCT simulations. The final RMSD and $R^2$ values are
then obtained by averaging over the entire test set with
$N_{\text{test}}=98$, see Table \ref{tab:Set1_perfomance_measures}.\\

\begin{table}[ht]
\centering
\begin{tabular}[t]{l|cc|cc}
\hline
\hline
DTD model&${\rm RMSD}_{\rm NN}$&${R^2}_{\rm NN}$&${\rm RMSD}_{\rm QCT}$&${R^2}_{\rm QCT}$\\
\hline
F-NN&0.0007&0.9996&0.0014&0.9982\\
K-NN&0.0013&0.9984&0.0014&0.9981\\
G-NN&0.0009&0.9994&0.0010&0.9991\\
\hline
\hline
\end{tabular}
\caption{Performance measures of the F-, K- and G-based models (F-NN,
  K-NN and G-NN) trained and evaluated on Set1. The number of
  significant digits being reported is based on the findings of Table~S2.}
\label{tab:Set1_perfomance_measures}
\end{table}%

\begin{figure}[h!]
\begin{center}
\includegraphics[width=0.99\textwidth]{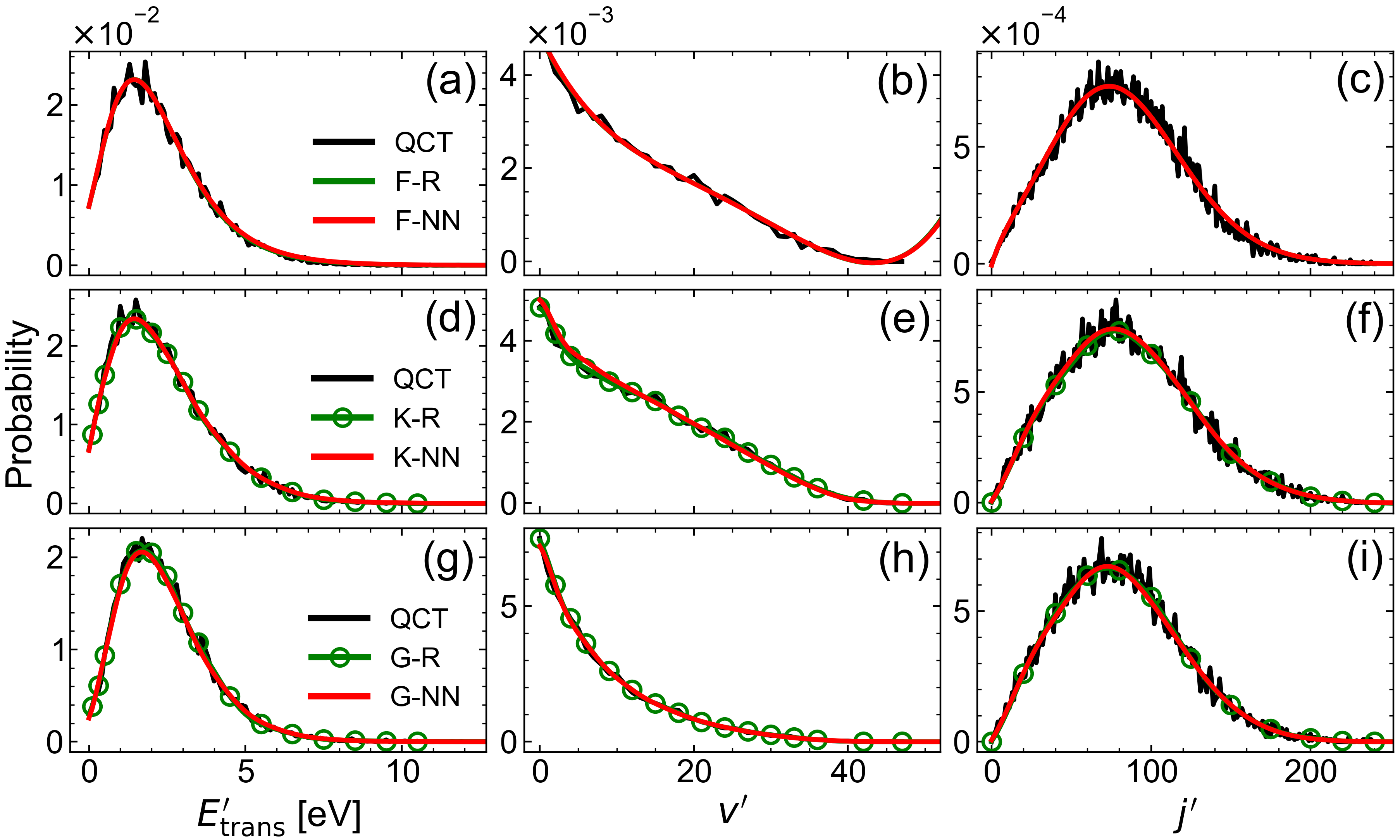}
\caption{Product state distributions obtained by explicit QCT
  simulations (QCT) as well as the corresponding references (-R) and
  predictions (-NN) from the (a-c) F-based (F-R, F-NN), (d-f) K-based
  (K-R, K-NN) and (g-i) G-based approaches (G-R, G-NN). Also, the
  amplitudes to construct the reference RKHS-based representations in
  the K- and G-based approaches are displayed (circles). The data sets
  considered here are from the test set of Set1 and result in
  predictions that are characterized by an $R^2_{\rm NN}$ value
  closest to the mean $R^2_{\rm NN}$ value as evaluated over the test
  set: (a-c) $\bm{T}=($9500 K, 16000 K, 16000 K),
  RMSD$_{\text{NN}}=0.0005$, $R^2_{\text{NN}}=0.9996$, (d-f)
  $\bm{T}=($10250 K, 19250 K, 19250 K), RMSD$_{\text{NN}}=0.0013$,
  $R^2_{\text{NN}}=0.9984$, (g-i) $\bm{T}=($12000 K, 9750 K, 9750 K),
  RMSD$_{\text{NN}}=0.0009$, $R^2_{\text{NN}}=0.9994$.}
\label{fig:fig5}
\end{center}
\end{figure}

\noindent
For three different data sets from the test set, the results from
explicit QCT simulations, the NN predictions, and the reference
representation of the corresponding approach are shown in Figure
\ref{fig:fig5}. These results are representative of NN predictions for
data from the test set for each of the three approaches as they are
characterized by an $R^2_{\rm NN}$ value closest to the mean $R^2_{\rm
  NN}$ value as evaluated over the test set. Figures
S1 and
S2 show the predictions that are
characterized by the highest (``accurate'' prediction) and lowest
(``inaccurate'' prediction) $R^2_{\rm NN}$ value in the test set,
respectively.\\

\noindent
In general, all three approaches are very accurate,
as their predictions closely match the corresponding reference
representations. Closer inspection of Figure \ref{fig:fig5} reveals
that for these particular examples the K-based approach appears to
yield slightly less accurate predictions than the other two
approaches (see, for example Figure \ref{fig:fig5}f).  The RMSD$_{\rm
  NN}$ values are 0.0007, 0.0013 and 0.0009 for the F-, K-, and
G-based models respectively, and the corresponding $R^2_{\rm NN}$
values are 0.9996, 0.9984 and 0.9994. These performance measures
indicate that the F-based approach yields the most accurate
predictions, followed by the G- and K-based approaches.\\

\begin{figure}[h!]
\begin{center}
\includegraphics[width=0.99\textwidth]{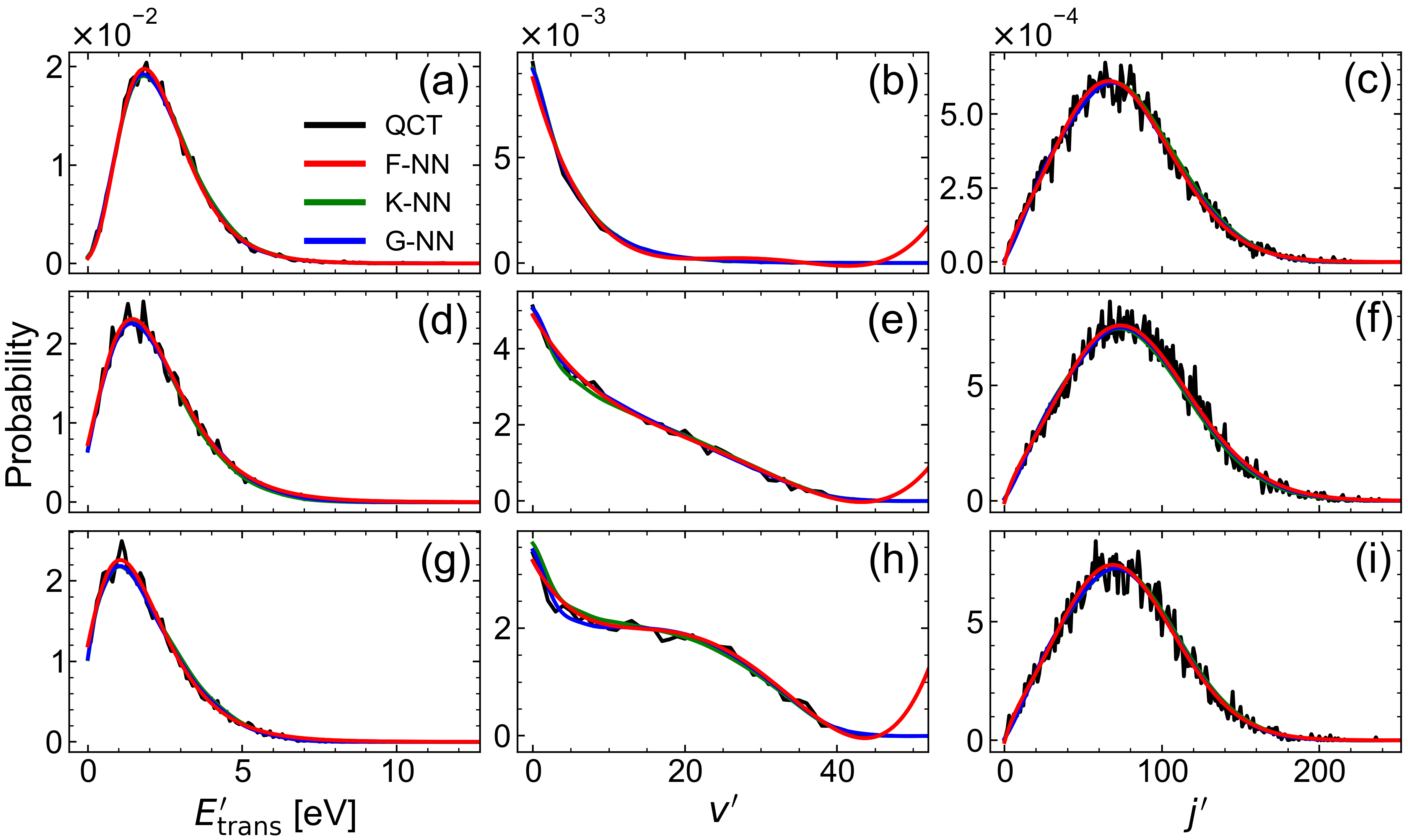}
\caption{Product state distributions obtained by explicit QCT
  simulations (QCT) as well as the corresponding model predictions
  obtained in the (a-c) F-based (F-NN), (d-f) K-based (K-NN) and (g-i)
  G-based approaches (G-NN). The data sets considered here are from
  the test set of Set1: (a-c) $\bm{T}=$ (12500 K, 5750 K, 5750 K) ,
  (d-f) $\bm{T}=$ (9500 K, 16000 K, 16000 K), (g-i) $\bm{T}=$ (5750 K,
  19250~K, 19250 K).}
\label{fig:fig6}
\end{center}
\end{figure}

\noindent
{\it 2.  Quality of the F-, K-, and G-Based Model Predictions:} For
the product state distributions the prediction accuracy depends on the
accuracy of the NN and the accuracy with which the representations
approximate them. Figure \ref{fig:fig6} compares the final model
predictions from the three approaches. The examples illustrate the
variety of product state distributions in Set1. It is found that
despite the appreciable variation in shapes (in particular for
$P(v')$) all three models correctly describe the product state
distributions. Distributions $P(v')$ are not well represented as a
single equilibrium distribution which is typical for vibrational
states at high temperatures.\cite{boyd:2015,schwartz:2018}\\

\noindent
A quantitative measure for the performance of the three models are the
RMSD and $R^2$ values calculated by comparing the locally averaged
($n_{\text{max},E_{\rm trans}'} = 3$, $n_{\text{max},v'} = 2$,
$n_{\text{max},j'} = 10$) QCT data and the model predictions following
a similar procedure as for RMSD$_{\rm NN}$ and $R^2_{\rm NN}$ which
will be called RMSD$_{\rm QCT}$ and $R^2_{\rm QCT}$, respectively. For
Set1 these performances are reported in Table
\ref{tab:Set1_perfomance_measures}. Again, all models are of high
quality with the G-based approach performing best. The somewhat lower
quality of the F-based approach when compared to the G-based approach
can largely be attributed to the fits of the product state
distributions. The representation of the F-based approach leads to
differences, in particular for $P(v')$ (e.g., deviations for small and
high $v'$ or extra undulations in Figure \ref{fig:fig6}b). However,
the deviations in the F-based approach observed for high $v'$ are only
partially relevant, as the accessible vibrational and rotational state
space is finite in practice, here $v'_{\rm max}=47$, $j'_{\rm
  max}=240$. Since state space is limited, extrapolation is not always
required. Considering the K- and G-based approaches, the reference
representations describing product distributions are nearly identical
and reproduce the QCT data very closely. Hence, the lower accuracy of
the final model in the K-based approach when compared to the G-based
approach can largely be attributed to its lower NN prediction
accuracy. \\

\noindent 
For the F-based approach, finding an optimal set of model functions
(Eqs. \ref{eq:func_E_r} to \ref{eq:func_j_p}) specific to the system
at hand is expected to be a difficult task. Such parametric models for
nonequilibrium conditions are still a current topic of
research.\cite{schwartz:2018, singh1:2019, singh2:2019} To highlight
the performance of different models, a parametric model for transient
vibrational and rotational state distributions based on surprisal
analysis\cite{schwartz:2018} was applied to Set1. $P(v')$ and $P(j')$
distributions for two different sets of temperatures
(($T_{\text{trans}} = 20000$ K, $T_{\text{rovib}} = 5000$ K) and
($T_{\text{trans}} = 5500$ K, $T_{\text{rovib}} = 20000$ K)) from QCT
simulation are modelled following the parametrization of
Ref. \citen{schwartz:2018} (see Figure~S3). While for the first set of temperatures
(translationally hot and rovibrationally moderately hot) the QCT
results for $P(v')$ and $P(j')$ are closely matched by the model, for
the second set (translationally moderately hot and rovibrationally
hot) both distributions are insufficiently described by the model, in
particular for $P(v')$ which is consistent with
Ref. \citen{schwartz:2018}. The fact that the shape of the $P(v')$
appears to vary more widely for different $\bm{T}$ compared to
$P(E_{\rm trans}')$ and $P(j')$ could be a partial explanation of why
developing a universally valid parametric model for $P(v')$ is more
challenging. It should be emphasised that comparison between
predictions based on the model and explicit QCT simulations is
mandatory to validate the model function used.\\

\subsection{Sensitivity of Performance to Feature Selection}
As in all machine learning tasks, feature selection for representing
the raw data is crucial for the complexity and prediction accuracy of
the resulting NN-based model.\cite{goodfellow:2016, bengio:2012} Here,
the main difference between the three approaches are the features that
represent reactant and product state distributions and which serve as
input/output of the NNs. Hence, any difference in the NN prediction
accuracy is due to the features used (i.e. the
``featurization''). Here, the features are fitting parameters (F-based),
kernel coefficients (K-based), and amplitudes (G-based) and together they
constitute a feature vector. Hence, a good featurization allowing for
an accurate NN to be trained is characterized by the fact that
similarly shaped distributions are described by similar feature
vectors.\cite{faber:2015} Here, ``similarity'' is measured by an
appropriate metric, such as an Euclidean norm.\\

\noindent
For the F-based approach, the choice of model functions (see
Eqs. \ref{eq:func_E_r} to \ref{eq:func_j_p}) turned out to
yield a satisfactory featurization. Conversely, for the K-based
approach it was necessary to increase the regularization rate
$\lambda$ and averaging over more neighbouring data points which in
essence smooths out sharp variations in the kernel coefficients
between neighboring data sets in temperature space. In the G-based
approach, accurate NN predictions were obtained through local
averaging because the amplitudes are the features.\\

\noindent
For the F-based approach the dependence of NN performance on feature
selection was explicitly explored by choosing an alternative
parametrization for
\begin{equation}
\tilde{P}(v') = g_{1}\exp(-(\ln(2g_{2}(v'-g_{3})/g_{4}+1)/g_{2})^2) +
g_{5}\exp(-(\ln(2g_{6}(v'-g_{7})/g_{8}+1)/g_{6})^2),
\label{eq:func_v_p_alternative}
\end{equation}
where $\bm{g}=(g_{1},...,g_{8})$ are the corresponding fitting
parameters. The resulting fit (see Figure~S4) to the QCT data
demonstrates that Eq. \ref{eq:func_v_p_alternative} yields a better
fit than Eq. \ref{eq:func_v_p}. However, training the corresponding NN
turned out to be difficult and the resulting NN predictions were
highly inaccurate (see below).\\

\noindent
Figure \ref{fig:pv_features} illustrates these points for $P(v')$ for
three combinations of simulation temperatures with 1)
$T_{\text{trans}} \sim T_{\text{rovib}}$ ($T_{\text{trans}} = 5000$ K,
$T_{\text{rovib}} = 5000,5250,5500,5750$ K; black) 2)
$T_{\text{trans}} < T_{\text{rovib}}$ ($T_{\text{trans}} = 5000$ K,
$T_{\text{rovib}} = 10000,10250,10500,10750$ K; red), and 3)
$T_{\text{trans}} > T_{\text{rovib}}$ ($T_{\text{trans}} = 12000$ K,
$T_{\text{rovib}} = 5000,5250,5500,5750$ K; green). As Figure
\ref{fig:pv_features}a demonstrates, the shapes of all $P(v')$ are
comparable and for each color there are four largely overlapping
distributions which can not be separated because the differences in
$T_{\text{rovib}}$ are too small. Using the F-based approach with
Eq. \ref{eq:func_v_p} for $P(v')$ yields parameter values that are
clustered (Figure \ref{fig:pv_features}b) for the black, red, and
green $P(v')$, respectively. For such input a robust NN can be
trained. Conversely, using Eq. \ref{eq:func_v_p_alternative}, even the
fitting parameters for one set of $P(v')$ spread considerably
and mix with those from $P(v')$ of other temperature combinations, see
Figure \ref{fig:pv_features}c. Thus, similarity in the shape of
$P(v')$ does not translate into similarity of the fitting parameters
used for the featurization. This is an unfavourable situation for
training a NN which compromises the prediction ability of such an
F-based model. The F-based model trained with Eq. \ref{eq:func_v_p}
yields an accurate prediction (${\rm RMSD}_{\rm NN}=0.0007, R^2_{\rm
  NN} = 0.9996$) whereas the one trained with
Eq. \ref{eq:func_v_p_alternative} fails to predict $P(v')$ (RMSD$_{\rm
  NN}=0.0348$, $R^2_{\rm NN}=-10.6133$; i.e., this model is worse than
a baseline model with $R^2 =0$). Similarly, a K-based approach can lead
to considerable spread of the kernel coefficients (Figure
\ref{fig:pv_features}d) which is not observed for the amplitudes in
the G-based approach (Figure \ref{fig:pv_features}e). Such differences
in the featurization leads to differences in the NN prediction
accuracies.\\

\begin{figure}[h!]
\begin{center}
\includegraphics[width=0.99\textwidth]{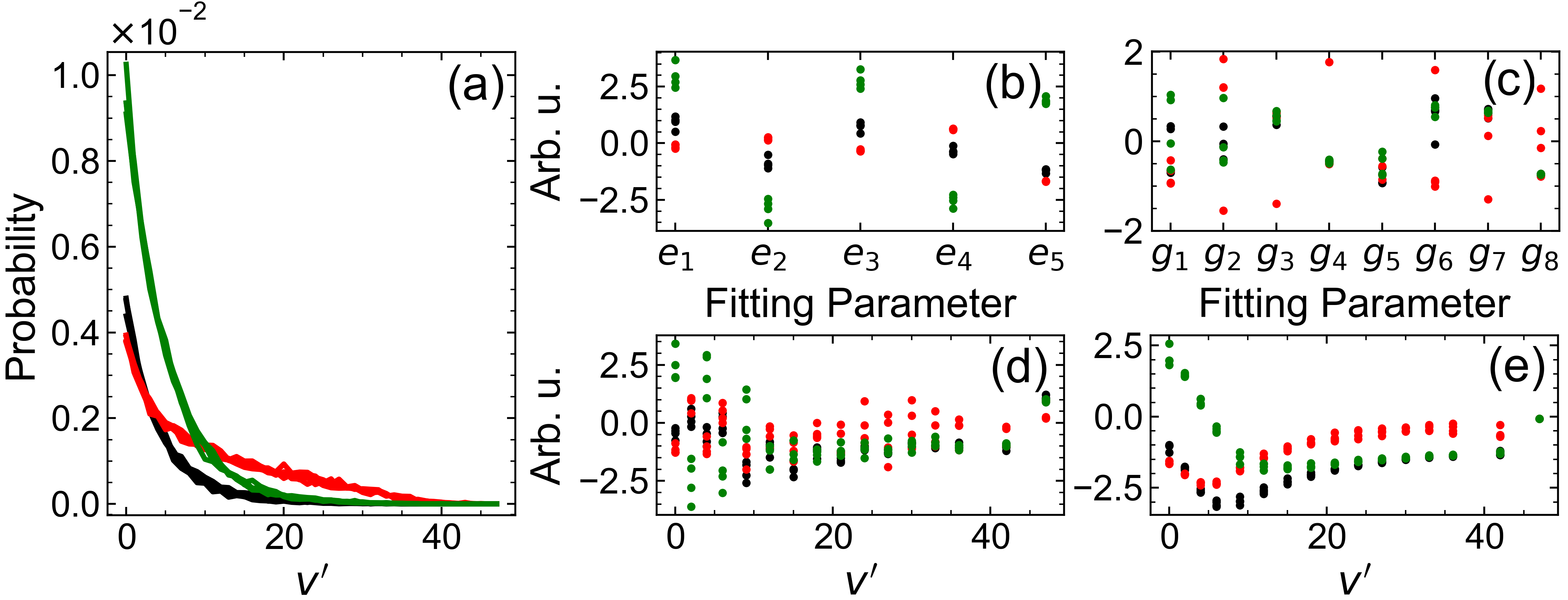}
\caption{Comparison of the featurization for three different groups of
  similarly shaped $P(v')$ (black, red and green). Panel a: the
  distributions as obtained from QCT simulations (fluctuating
  lines). Panels b and c: Fitting Parameters for the F-based approach
  with model functions from Eqs. \ref{eq:func_v_p} and
  \ref{eq:func_v_p_alternative}; panel d: kernel coefficients for the
  K-based approach and panel e: amplitudes for G-based approach. The
  temperatures are ($T_{\text{trans}} = 5000$ K, $T_{\text{rovib}} =
  5000,5250,5500,5750$ K; black), ($T_{\text{trans}} = 5000$ K,
  $T_{\text{rovib}} = 10000,10250,10500,10750$ K; red), and
  ($T_{\text{trans}} = 12000$ K, $T_{\text{rovib}} =
  5000,5250,5500,5750$ K; green). The quality of all fits in panel a
  is as good as in Figure \ref{fig:fig6} and all features are
  standardized.}
\label{fig:pv_features}
\end{center}
\end{figure}

\noindent 
Another difference between the NNs in the three approaches is the fact
that prediction errors in the features translate into errors in the
corresponding predicted product state distributions in different
ways. For the G-based approach, an error in the predicted features
directly translates into an error in the predicted product state
distributions. This is not the case for the F- or K-based
approaches. As an example, the model functions for the product state
distributions in an F-based approach depend nonlinearly on the fitting
parameters (features). Hence, small errors in the NN predictions can
lead to large errors in the predicted distributions. This is
problematic, as the NNs are trained on a loss function that measures
the errors in feature space, whereas one is rather interested in the
quality of the predicted product state distributions. By using a loss
function that depends on errors in the predicted product distributions
this problem can be avoided. In the K-based approach this is partially
resolved by the choice of a Gaussian kernel where the hyperparameter
$\sigma$ is assigned according to the procedure described in Section
\ref{K-based method}. This results in a local kernel with kernel
coefficients largely determined by the amplitude at the corresponding
grid points and its nearest neighbours.\cite{bengio:2006}
Consequently, errors in the predicted kernel coefficients are also
restricted to impact the predicted model function locally, similar to
the G-based approach.\\

\subsection{Computational Cost and Generalizability}
To compare the computational cost of the final models in the F-, K-
and G-based approaches, the evaluation times of the final models for
1000 randomly selected data sets (from Set1) are considered. Here, a
single evaluation is defined as a prediction of the product
($E_{\text{trans}}',v',j'$) distributions at 201, 48 and 241 evenly
spaced points in the interval between $E_{\text{trans}}' = 0-20$ eV,
$v' = 0-47$ and $j' = 0-240$, respectively, given the reactant state
distributions. The evaluation times on a 1.8 GHz Intel Core i7-10510U
CPU are $(9.0 \pm 0.1)$ s, $(29.0 \pm 0.3)$ s, and $(28.9 \pm 0.3)$ s
for the F-, K- and G-based models, respectively. For the F-based
approach the evaluation time is 3 times faster compared to the two
other methods and is dominated by fitting the reactant state
distributions to Eqs. \ref{eq:func_E_r} to \ref{eq:func_j_r} whereas
for the K- and G-based approaches the evaluation time is dominated by
the evaluation of the RKHS-based representations of the product
distributions given the predicted kernel coefficients or amplitudes,
respectively. This may be further improved for the K- and G-based
approaches by using a computationally efficient kernel
toolkit.\cite{MM.RKHS:2017}\\

\noindent 
In terms of generality and transferability, an F-based model can {\it
  not} be easily generalized to distributions with widely different
shapes emanating from the QCT simulations. New optimal model
functions, also suitable for training a NN would need to be found for
every single system. Conversely, with the K- and G-based approaches
all desired features of the distributions can be captured by an
appropriate choice of reproducing kernel and grid, requiring
fine-tuning of the corresponding hyperparameters. Compared to the
K-based approach, a G-based model only requires tuning of the
corresponding hyperparameters for the product state distributions. In
addition, for a G-based model it is also possible to use a linear
interpolation instead of a RKHS if one is not concerned with
extrapolation. Then, the grid for the product state distributions
needs to be chosen sufficiently dense suitable for linear
interpolation at the cost of an increased number of NN
parameters.\\

\subsection{Grid-Based Models for $T_{\text{vib}} \neq T_{\text{rot}}$}
As vibrational relaxation is often slow in hypersonic flow, assuming
$T_{\text{vib}} = T_{\text{rot}}$ is often not a good
approximation.\cite{boyd:2015,bender:2015} Therefore, the G-based
approach is extended to and tested for the case of $T_{\rm vib} \neq
T_{\rm rot}$ using Set2. Restricting this to the G-based approach is
motivated by the fact that it performed best so far, both in terms of
final model accuracy and practicability.\\

\begin{figure}[htb!]
\begin{center}
\includegraphics[width=0.99\textwidth]{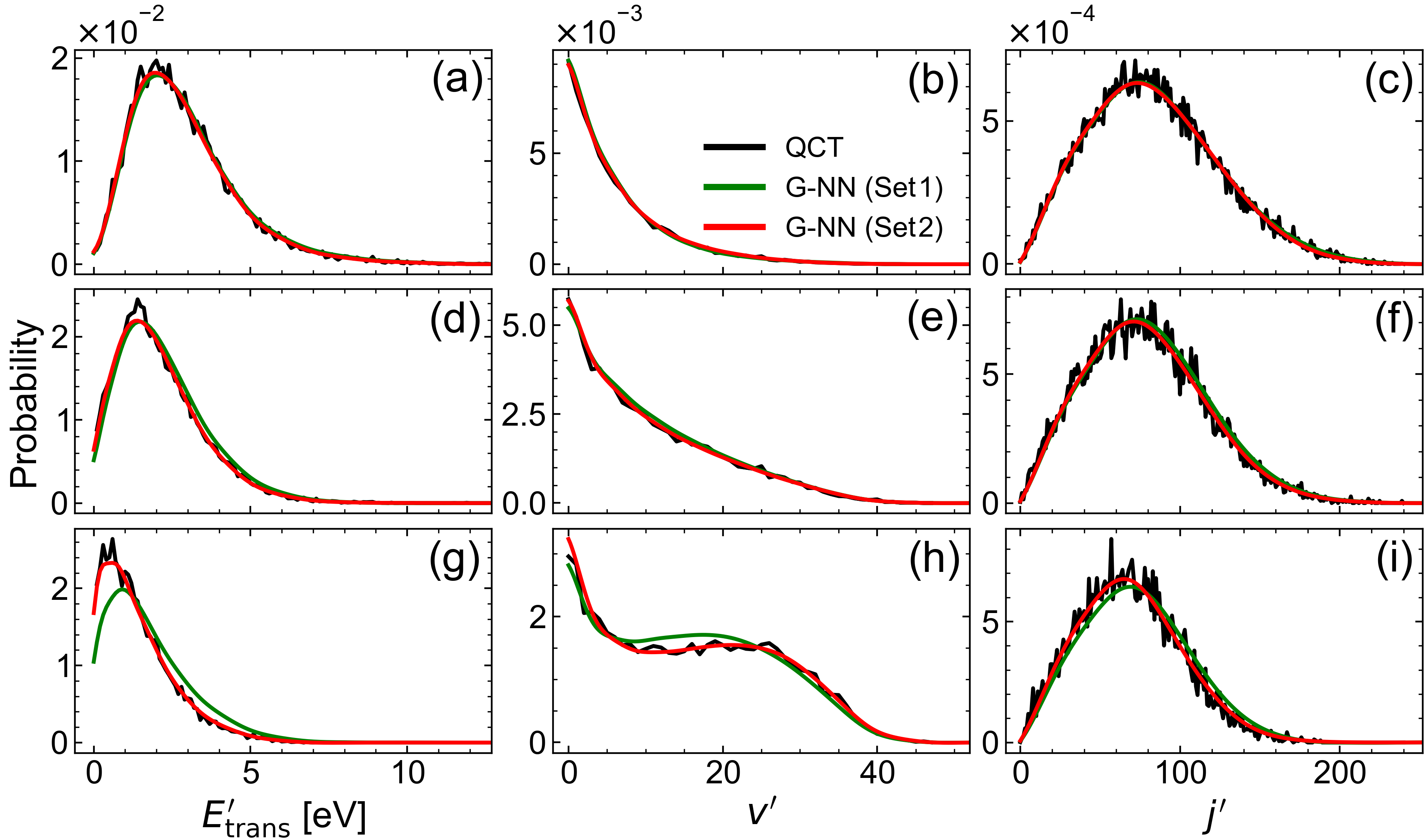}
\caption{Performance of the G-based approach on Set2 from training on
  Set1 (green) or Set2 (red). Product state distributions from
  explicit QCT simulations (QCT) compared with model predictions from
  the G-based approach by training on Set1 (G-NN (Set1)) and Set2
  (G-NN (Set2)) for: (a-c) $\bm{T}$ = (20000 K, 7000 K, 5000 K) ,
  (d-f) $\bm{T}$ = (10000 K, 14000 K, 9000 K), (g-i) $\bm{T}$ = (5000
  K, 20000 K, 8000 K).}
\label{fig:tunequal}
\end{center}
\end{figure}

\noindent
First, predictions for Set2 were made based on the G-based model
(G-NN) trained on Set1 ($T_{\text{vib}} = T_{\text{rot}}$) and
compared with QCT data (QCT), see Figure \ref{fig:tunequal}. The
accuracy of this G-based model deteriorates (see Table
S3 for all performance measures) as the
difference between $T_{\text{vib}}$ and $T_{\text{rot}}$ increases
(green lines in Figure \ref{fig:tunequal}). Consequently, a new
G-based model was trained and evaluated on Set2. The resulting model
predictions (red lines in Figure \ref{fig:tunequal}) are very
accurate, close to the level of accuracy of the G-based model trained
and evaluated on Set1. Thus, the G-based approach performs equally
well for $T_{\text{vib}} = T_{\text{rot}}$ and $T_{\text{vib}} \neq
T_{\text{rot}}$.\\

\noindent
In an attempt to further improve the G-based approach, three
alternative types of input were considered. They are all based on
reducing the number of input which not only decreases computational
cost, but also removes redundant features which can improve prediction
accuracy.\cite{chandra:2014} Again, continuous distributions were
obtained from an RKHS of the discrete predictions. The first
(G$'$-based) model used values of the reactant state distributions at
a fixed but reduced number of grid points compared with the G-based
model used so far (see Table S1). Next, a model
using only the three temperatures characterizing the reactant state
distributions $\bm{T}$ as input (G($\bm{T}$)-based model) is
considered. This is meaningful because the value of $\bm{T}$ entirely
specifies the equilibrium reactant state distributions. A third model
used averages $\bm{\mu}=(\mu_{E_{\rm trans}},\mu_{v},\mu_{j})$ of the
reactant state distributions as input (G($\bm{\mu}$)-based model). For
generality, these models will be trained and evaluated on Set2 and all
performance measures are summarized in Table
S3.\\

\noindent
A G$'$-based approach using two grid points per reactant state
distribution ($E_{\text{trans}}=0.3,3.5$ eV; $v=2,12$; $j=30,150$)
still allows for accurate predictions of the product state
distributions. The location of these grid points is largely arbitrary,
but they should be sufficiently spaced to provide information about
the distribution at different locations. However, reducing to a single
grid point per reactant state distribution ($E_{\text{trans}}=0.6$ eV,
$v=6$, $j=60$) leads to a significant drop in the prediction
accuracy. The fact that 2 grid points per reactant state distribution
are required for accurate predictions can mainly be attributed to the
presence of noise in the distributions arising from finite sample
statistics (see Section IV in SI
for further clarification). The resulting predictions for the above
mentioned models for selected data sets from the test set are
displayed in Figure~S6 in the SI.\\

\noindent
For the G($\bm{T}$)-based model the performance is close to the
original G-based model trained and evaluated on Set2. This is
expected, as $\bm{T}$ entirely specifies the equilibrium reactant
state distributions and allows a NN to predict corresponding product
state distributions. Finally, providing the mean $\mu$ of each of the
reactant state distributions as input in a G($\bm{\mu}$)-based model
also leads to highly accurate predictions. This can be explained by
the fact, that the mean values $\bm{\mu}$ of the reactant equilibrium
distributions are uniquely linked to the corresponding set of
temperatures $\bm{T}$.\\

\noindent
The results of this Section highlight that all three {\it variants} of
the G-based model yield similarly high levels of accuracy as the
G-based approach, which makes them preferable as they are
computationally less expensive. These results may be specific to
reactant state distributions that can be uniquely specified by a
single parameter, such as a temperature or its mean value $\mu$. To
explore this and to further demonstrate the generality of the G-based
approach and its variants, a more diverse dataset for nonequilibrium
conditions (Set3) was finally considered, for which the reactant state
distributions are characterized by multiple sets of temperatures
$\bm{T}$.\\

\subsection{Grid-based Models for Nonequilibrium Product State Distributions}
As a final application, nonequilibrium DTD models are constructed for
Set3 which was generated by means of a weighted sum (see
Eq. \ref{eq:multit}) using Set2 (see Section
\ref{Generating_noneq_Datasets}) with $N \in [2,3]$, and $w_{n} \in
    [1,2]$. Training and validation sets of variable sizes were
    considered, whereas $N_{\text{test}}=125$ throughout. First, a
    G-based model was trained on Set3 with
    $N_{\text{train}}+N_{\text{valid}}=5000$. Again, all performance
    measures are summarized in Table
    S3.\\

\begin{figure}[th!]
\begin{center}
\includegraphics[width=0.99\textwidth]{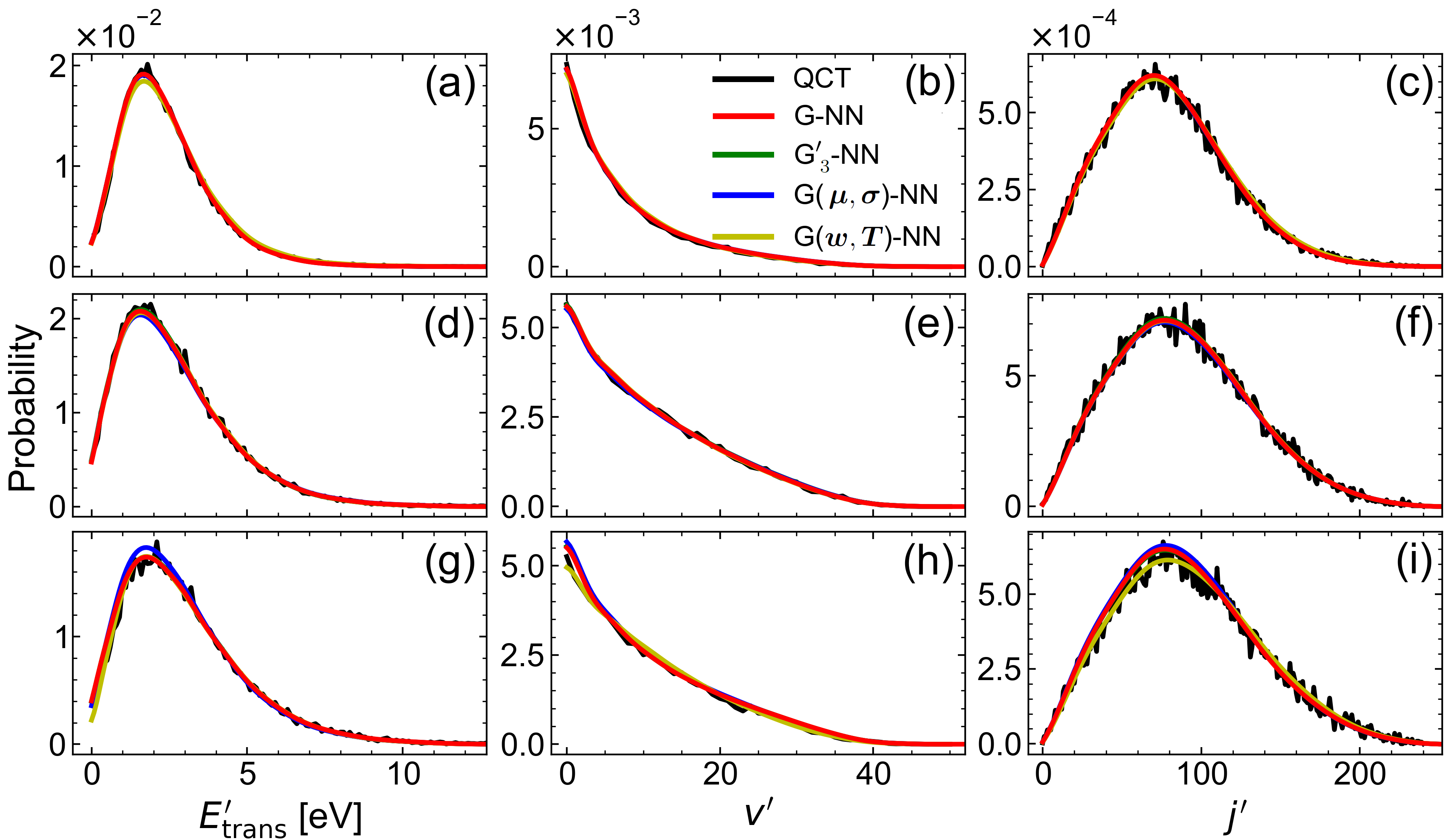}
\caption{Performance of G-based models and variants trained and evaluated on
  Set3. Product state distributions obtained from explicit QCT
  simulations (QCT), together with the predictions from the G- (G-NN),
  G$'_{3}$- (G$'_{3}$-NN), G($\bm{\mu},\bm{\sigma}$)-
  (G($\bm{\mu},\bm{\sigma}$)-NN) and G($\bm{w},\bm{T}$)-based
  (G($\bm{w},\bm{T}$)-NN) models trained on Set3. G$'_{3}$-NN uses 3
  grid points per reactant state distribution (see text). The data
  sets considered here are from the test set of Set3. In particular,
  the predictions for these three data sets are characterized by (a-c)
  a $R^2_{\rm QCT}$ value closest to the mean $R^2_{\rm QCT}$ value as
  evaluated over the entire test set, as well as (d-f) the largest and
  (g-i) smallest $R^2_{\rm QCT}$ value in the test set,
  respectively. The normalized weights $w_{n}/w_{\rm tot}$ and sets of
  temperatures $\bm{T}$ characterizing the data sets displayed here
  are given in Table S4.}
\label{fig:probmodel_differentinputs_R2}
\end{center}
\end{figure}

\noindent
The predictions of this G-based model for three different data sets
from the test set are shown in Figure
\ref{fig:probmodel_differentinputs_R2}. In particular, the predictions
for these three data sets are characterized by a $R^2_{\rm QCT}$ value
closest to the mean $R^2_{\rm QCT}$ value as evaluated over the entire
test set, as well as the highest and lowest $R^2_{\rm QCT}$ value in
the test set, respectively. Once again, the G-based approach gives a
very accurate DTD model, close to the G-based model trained and
evaluated on Set2. However, the predictions of the G-based model for
Set3 have a larger variance compared to the G-based model for Set2. To
assess the influence of the training and validation set size on the
prediction accuracy, a learning curve was computed (Figure~S7). The NN prediction accuracy does
not significantly increase when $N_{\text{train}}+N_{\text{valid}}$
was increased from 5000 to 30000 in increments of 5000. Hence, this
justifies training and validating the variants of the G-based model
only on $N_{\text{train}}+N_{\text{valid}} = 5000$.\\

\noindent
In an attempt to further improve and reduce this model for Set3, the
dependence on different amount of input information (as was done for
Set2) was tested again. Accurate predictions are still possible with
amplitudes of $P(E_{\text{trans}}), P(v), P(j)$ at three different
grid points ($E_{\text{trans}}=0.3,1.5,3.5$ eV; $v=2, 6, 12$; $j = 30,
60, 150$), but the prediction accuracy decreases when reducing this to
two grid points ($E_{\text{trans}} = 0.3, 3.5$ eV; $v = 2,12$; $j =
30, 150$). Again it is found that with a G$'$-based approach the
number of grid points characterizing the reactant state distributions
can be significantly reduced compared to the grids in Table
S1. This suggests that the possibility to reduce
the number of input features (amplitudes) in such G-based models is a
generic property which can be systematically explored.\\

\noindent
Providing the mean $\mu$ and standard deviation $\sigma$ for each
reactant state distribution ($E_{\text{trans}}$,$v$,$j$) in a
G($\bm{\mu},\bm{\sigma}$)-based approach also yields a good model
whereas omitting the standard deviations $\bm{\sigma}$ as input
information results in a G($\bm{\mu}$)-based model with a
significantly lower prediction accuracy. This should be compared with
the G($\bm{\mu}$)-based model for Set2 which yielded accurate
predictions. The aforementioned differences of the G$'$- and
G($\bm{\mu}$)-based models for Set2 and Set3 can be attributed to the
fact that the reactant state distributions in Set3 show more diverse
shapes compared to Set2, which makes it necessary to provide
additional information to maintain a high prediction accuracy. In
particular, reactant state distributions in Set3 are nonequilibrium
distributions and consequently can {\it not} be uniquely specified by
a single parameter, such as a temperature or its mean value $\mu$, as
was the case in Set2. Rather, the G($\bm{\mu},\bm{\sigma}$)-based
approach for Set3 showed that the set of reactant state distributions
is characterized by specifying ($\bm{\mu},\bm{\sigma}$).\\

\noindent
Keeping this in mind, extending the G($\bm{T}$)-based approach to Set3
can be achieved by providing the sets of temperatures $\bm{T}$ from
which the particular reactant state distributions of Set3 were
generated, together with the set of weights $\bm{w}$ with which these
contributed (see Section \ref{Generating_noneq_Datasets}). This
results in a G($\bm{w},\bm{T}$)-based model. To always guarantee the
same number of NN input being specified, as expected by the NN used in
this work (see Section \ref{Neural Network}), zero padding was
used. Such a G($\bm{w},\bm{T}$)-based model for Set3 leads to accurate
predictions. The predictions of the G$'$-, G($\bm{\mu},\bm{\sigma}$)-
and G($\bm{w},\bm{T}$)-based model for selected data sets from the
test set are also reported in Figure
\ref{fig:probmodel_differentinputs_R2}.\\

\noindent 
Interestingly, the G-, G$'$- and G($\bm{\mu},\bm{\sigma}$)-based
models trained on Set3 can accurately predict the product state
distributions when given the reactant state distributions from
Set2. The performance measures (see Table
S3) were calculated by considering the
subset of 158 data sets of Set2 from which the test set of Set3 was
generated. Even though these models were trained on reactant state
distributions given as a linear combination of two to three
equilibrium distributions, they can accurately generalize to
equilibrium reactant state distributions. This is not the case for the
G($\bm{w},\bm{T}$)-based model, which yields unreliable predictions
when applied to the reactant state distributions of Set2. This can be
attributed to the zero padding. In particular, such a model can not
generalize {\it at all} to reactant state distributions being composed
of more than three equilibrium distributions, as this requires more NN
input to be specified than there are input nodes. This point can be
addressed in the future by considering a different NN architecture,
allowing for a variable input size.\\

\begin{figure}[th!]
\begin{center}
\includegraphics[width=0.99\textwidth]{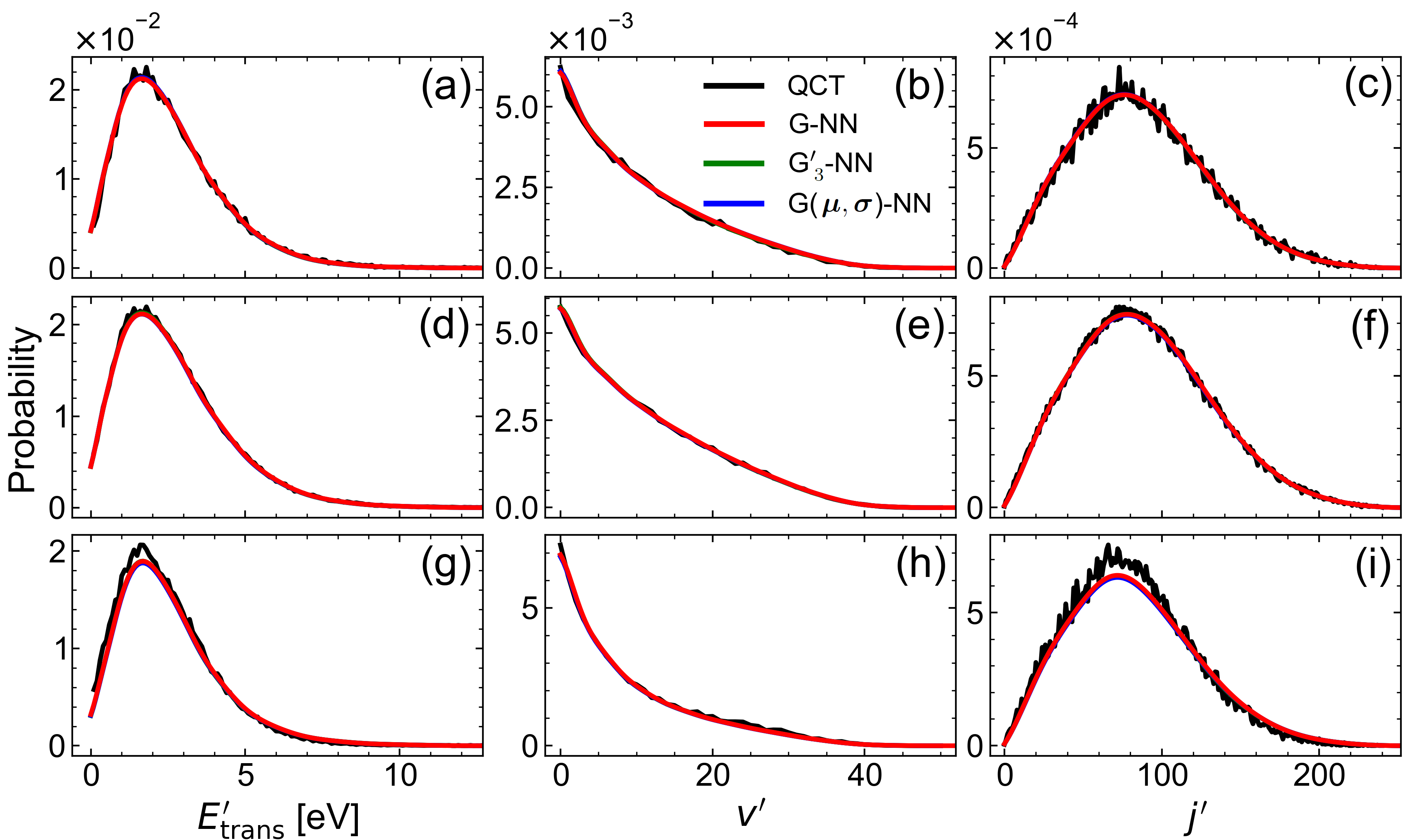}
\caption{Performance of G-based models and variants trained on Set3
  and evaluated on Set3A. Product state distributions obtained from
  explicit QCT simulations (QCT), together with the predictions from
  the G- (G-NN), G$'_{3}$- (G$'_{3}$-NN), G($\bm{\mu},\bm{\sigma}$)-
  (G($\bm{\mu},\bm{\sigma}$)-NN) and G($\bm{w},\bm{T}$)-based
  (G($\bm{w},\bm{T}$)-NN) models trained on Set3. G$'_{3}$-NN uses 3
  grid points per reactant state distribution (see text). The data
  sets considered here are from Set3A. In particular, the predictions
  for these three data sets are characterized by (a-c) a $R^2_{\rm
    QCT}$ value closest to the mean $R^2_{\rm QCT}$ value as evaluated
  over the entire Set3A, as well as (d-f) the largest and (g-i)
  smallest $R^2_{\rm QCT}$ value in Set3A, respectively. The
  normalized weights $w_{n}/w_{\rm tot}$ and sets of temperatures
  $\bm{T}$ characterizing the data sets displayed here are given in
  Table S5.}
\label{fig:probmodel_diverse_differentinputs_R2}
\end{center}
\end{figure}

\noindent
Conversely, generalizing to reactant state distributions being
composed of more than three equilibrium distributions is possible for
the G-, G$'$- and G($\bm{\mu},\bm{\sigma}$)-based models trained on
Set3. Specifically, a set of 125 reactant and product state
distributions given as a linear combination of $N \in [1-10]$
distributions from Set2 (i.e., from the subset of 158 data sets of
Set2 from which the test set of Set3 was generated) with integer
weights $w_{n} \in [1-100]$ (see Section
\ref{Generating_noneq_Datasets}) was generated, referred to as
Set3A. When applied to the reactant state distributions of Set3A,
these models still predict the corresponding product state
distributions with high accuracy, see Table
S3. The final DTD model predictions using
the G-based approach as well as its variants for such data sets is
shown in Figure
\ref{fig:probmodel_diverse_differentinputs_R2}. Consequently, as
discussed in Section \ref{Generating_noneq_Datasets}, these models are
also expected to generalize well to most nonequilibrium distributions
encountered in practice.\\

\section{Discussion and Conclusions}
The present work demonstrates that machine learning of product state
distributions from the corresponding reactant state distributions for
reactive atom + diatom collisions based on a NN (DTD model)
constitutes a promising alternative to a full but computationally very
demanding (or even unfeasible, e.g., for diatom + diatom type
collisions) treatment by means of explicit QCT simulations. For such
DTD models, only a subset of the state space of the reactant needs to
be sampled which drastically reduces the computational complexity of
the problem at hand.\cite{MM.nncs:2019} In particular, DTD models for
the N + O$_2$ $\rightarrow$ NO + O reaction were constructed following
three distinct (F-, K- and G-based) approaches for $T_{\rm vib} =
T_{\rm rot}$. Although all three approaches yield accurate predictions
for the product state distributions as judged from $R^2$ and RMSD
measures, the G-based approach performs best in terms of prediction
accuracy, generality and practical implementation. On the other hand
the F-based approach is computationally more efficient by a factor of
3 compared with the K- and G-based approaches. For the K- and G-based
approaches it is found that RMSD$_{\rm NN}$ and $R^2_{\rm NN}$ are
close to RMSD$_{\rm QCT}$ and $R^2_{\rm QCT}$, respectively. This is
different for the F-based approach, where RMSD$_{\rm NN}$ is smaller
than RMSD$_{\rm QCT}$ by a factor of 2 (similarly for $R^2$), see
Table \ref{tab:Set1_perfomance_measures}. This indicates that the
parametrizations used for the F-based model can still be improved. In
general, an F-based approach is feasible if a universally valid and
accurate parametrization for the distributions can be found, which
also allows for an accurate NN to be trained. However, finding such a
parametrization may not always be possible. Consequently, the G-based
approach is generally preferred.\\

\noindent
The G-based approach and its input-reduced variants (G$'$-,
G($\bm{\mu},\bm{\sigma}$)) were found to perform well, too, for
$T_{\rm vib} \neq T_{\rm rot}$ (Set2) and nonequilibrium reactant
state distributions (Set3 and Set3A). Consequently, the G$'$- and
G($\bm{\mu},\bm{\sigma}$)-based models are generally preferred over
the standard G-based model, as the reduced number of input lowers
their computational cost. Moreover, G-, G$'$- and
G($\bm{\mu},\bm{\sigma}$)-based models trained on Set3 are also
expected to generalize well to most realistic nonequilibrium
distributions. This is of particular relevance for applications in
hypersonics for which nonequilibrium effects are of importance.\\

\begin{figure}
\centering \includegraphics[width=3.5in]{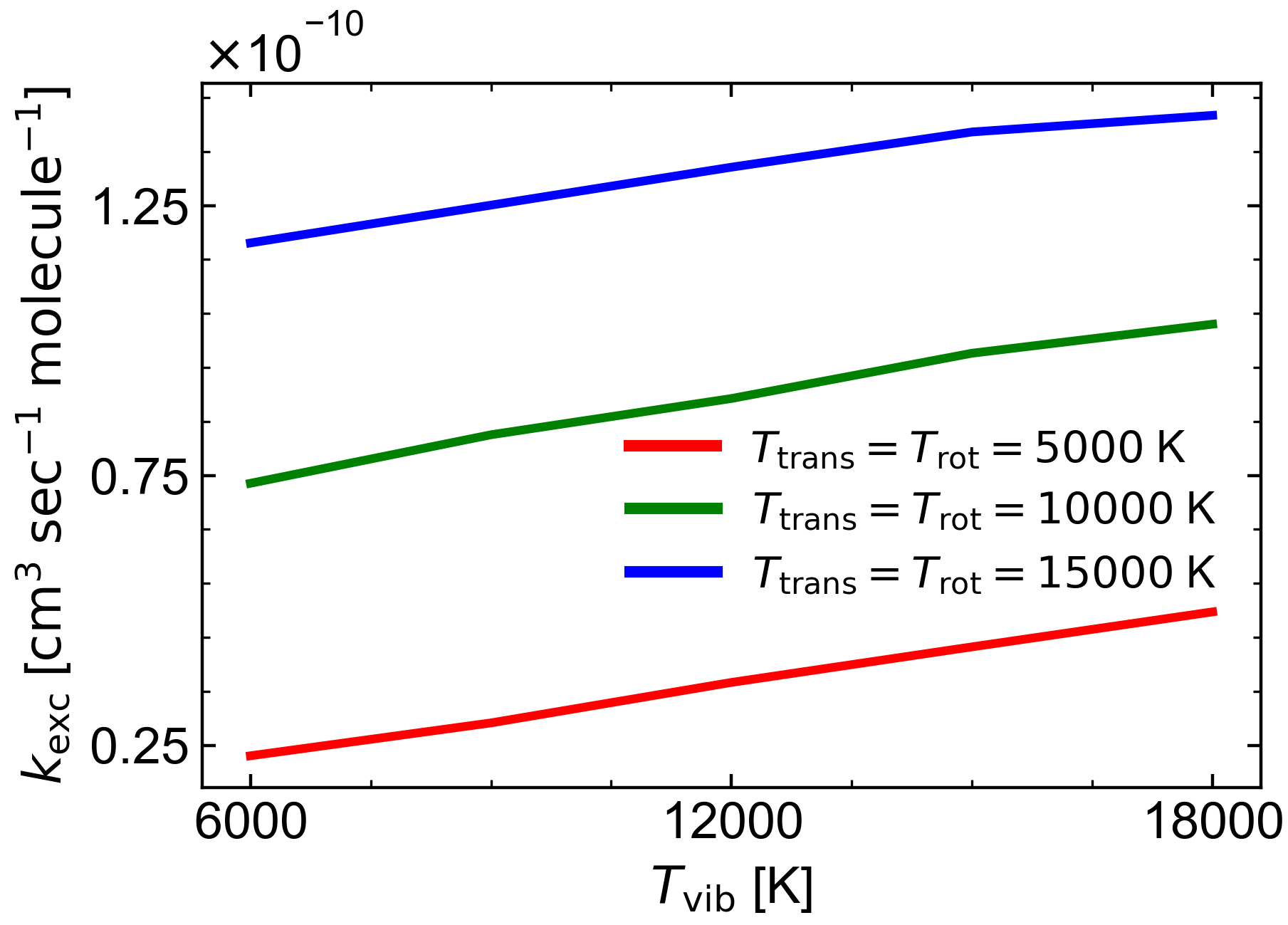}
\caption{Exchange reaction rates $k_{\rm exc} (\bm{T})$
  for N + O$_2$ $\rightarrow$ NO + O as a function of $T_{\rm vib}$, where
  $T_{\rm trans} = T_{\rm rot}$.}
\label{fig:fig10}
\end{figure}

\noindent
Therefore, it is also of interest to discuss the present findings in
the context of the methods traditionally employed in DSMC\cite{dsmc}
and CFD simulations for hypersonics. Continuum-level reaction rates
are required in a multi-temperature framework usually employed in CFD
solvers\cite{gnoffo1990upwind,wright1998data,nompelis2011implementation}. The
expression for the exchange reaction rates
\begin{equation}
k_{\rm exc} (\bm{T})  = \left(\cfrac{8 k_B T_{\rm trans}}{\pi \mu}
\right)^{1/2} \pi b_{max}^2 P_{\rm r},
\end{equation}
where $k_B$ is the Boltzmann constant, $\mu$ is the reduced mass of
the reactants. Here, $P_{\rm r}$ is the reaction probability, which
can be obtained in a computationally inexpensive way by integrating
one of the predicted product state distributions. While rates derived
in this manner are based on equilibrium distributions characterized by
$\bm{T}$, the vibrational population is nonequilibrium at high
temperatures ($T \geq 8000$ K).\cite{schwartz:2018} Nonequilibrium
effects are particularly relevant for diatomic dissociation because
high vibrational states have significantly increased probability for
dissociation. For instance, for the dissociation of N$_2$ (in N$_2$ +
N$_2$) studied in Ref. \citen{bender:2015}, at $T_{\rm trans} =T_{\rm
  rot} = 10000$ K, the dissociation probability to form N$_2$ + N + N
increases by a factor of 500, when $T_{\rm vib}$ is increased from
8000 K to 20000 K. Conversely, for the exchange reaction considered in
the present work, at $T_{\rm trans} = T_{\rm rot} = 10000$ K,
increasing $T_{\rm vib}$ from 5000 K to 18000 K results in an increase
in the reaction rate by only 40 \% (see Figure
\ref{fig:fig10}). Therefore due to the weaker dependence of the
exchange reaction probability on vibrational energy, a Boltzmann
distribution at $T_{\rm vib}$ may be sufficient for modeling exchange
rates. However, if necessary, the simple model for non-Boltzmann
distribution developed in Ref. \citen{singh1:2019} can be approximated
by a linear combination of Boltzmann distributions to include
non-Boltzmann effects in the reaction rates for exchange reactions as
well. The reactant state distributions can then expressed as a linear
combination of equilibrium distributions, as was done here for Set3,
from which $P_{\rm r}$ can be calculated. Furthermore, the average
vibrational energy change due to decomposition reactions, another key
input required in CFD, can also be obtained by taking an appropriate
moment of the product state distributions.\cite{singh2:2019}\\

\noindent
As an alternative to CFD for hypersonic flow, coarse grained Master
equations (ME) are being used for modeling chemical
kinetics.\cite{panesi2013rovibrational,magin2012coarse,andrienko2017state}
Here, several rovibrational states are lumped together in groups and
only the transition rates between these groups are required which
considerably speeds up such simulations. The accuracy of such an
approach directly depends on the criterion with which the groups are
generated,
though.\cite{macdonald2018_QCT,macdonald2018construction_DMS} A DTD
model as developed here constitutes a new framework for tracking the
time evolution of the population in each rovibrational state in a
computationally feasible manner. In the context of the present work
the DTD model can be repeatedly used for drawing reactant state
distributions at each time step for propagating the ME. This is
similar to sequential QCT proposed in
Refs. \cite{bruehl1988theoretical1,bruehl1988theoretical2} and DMS
method \cite{schwartzentruber:2018}, but computationally more
efficient because it \textit{avoids} explicit trajectory
calculations.\\

\noindent
The product state distributions predicted from the DTD models can also
be used for developing simple function-based, state-specific models
for exchange reactions in
DSMC.\cite{bird1976molecular,boyd2017nonequilibrium,singh1:2019,gimelshein2017modeling}
Such a model can be used within DSMC to estimate state-specific
exchange (forward and backward) reaction probabilities instead of the
total collision energy (TCE) \cite{bird1981monte} model. Furthermore,
DTD models also provide a QCT-, physics-based alternative to the
phenomenological Borgnakke-Larsen model\cite{borgnakke1975statistical}
which is currently employed to sample internal energy and
translational energy of products formed in exchange reactions.\\

\noindent
There is scope to further extend and improve the present methods. One
of them concerns the application of the G-based models to predict
product state distributions which can subsequently be used as reactant
state distributions for QCT simulations or DTD models. This way,
starting from a set of reactant state distributions transient
distributions can be obtained after a certain number of cycles. This
will be of particular relevance for applications in
hypersonics. Moreover, data construction schemes, such as constructing
nonequilibrium distributions as a linear combination of equilibrium
distributions, may prove useful for training DTD models from a small
set of reactant and product state distributions obtained from explicit
QCT simulations that generalize far beyond the conditions used for the
reactant state distributions.\\

\noindent
Overall, the present work establishes that NN-based models for
distribution-to-distribution learning can be developed based on
explicit trajectory-based data. This will apply to both, data
generated from QCT and quantum simulations if sufficiently converged
and complete data can be generated. More generally, the approach
presented in this work will also be applicable to situations in which
initial distributions are mapped on final distributions by means of a
deterministic algorithm such as molecular dynamics simulations. In the
future it may also be of interest to consider a fourth approach to DTD
learning based on a distribution regression network\cite{kou:2019}
promising a higher prediction accuracy with fewer NN parameters
compared to the approaches investigated in this work. Moreover, it may
also be interesting to explore the possibility for constructing a
``state-to-distribution'' model which would be intermediate between
the DTD model and the earlier STS model\cite{MM.nncs:2019}.\\

\section*{Data and Code Availability}
All data required to train the NNs has been made available on zenodo
\url{https://doi.org/QQQ/zenodo/QQQ} and the code for training the DTD
models is available at \url{https://github.com/MMunibas/DTD}.

\section{Acknowledgments}
This work was supported by the Swiss National Science Foundation
through grants 200021-117810, 200020-188724, the NCCR MUST, and the
University of Basel.

\bibliography{refs}
\end{document}

\title{Supplementary Information: Machine Learning for Observables:
  Reactant to Product State Distributions for Atom-Diatom Collisions}

\author{Julian Arnold, Debasish Koner, Silvan K\"aser}
\affiliation{Department of Chemistry, University of Basel,
  Klingelbergstrasse 80, CH-4056 Basel, Switzerland}

\author{Narendra Singh} \affiliation{Department of Mechanical
  Engineering, Stanford University, CA 94305 USA}

\author{Raymond J. Bemish} \affiliation{Air Force Research Laboratory,
  Space Vehicles Directorate, Kirtland AFB, New Mexico 87117, USA}

\author{Markus Meuwly}\email{m.meuwly@unibas.ch}
\affiliation{Department of Chemistry, University of Basel,
  Klingelbergstrasse 80, CH-4056 Basel, Switzerland}

\maketitle

\section{Grids}

\begin{table}[ht]
\centering
\begin{tabular}[t]{cccccc}
\hline
\hline
$E_{\text{trans}}$ [eV]&$v$&$j$&$E_{\text{trans}}'$ [eV]&$v'$&$j'$\\
\hline
0.0&0&0&0.1&0&0\\
0.2&2&15&0.3&2&20\\
0.4&4&30&0.5&4&40\\
0.6&6&45&1.0&6&60\\
1.0&9&60&1.5&9&80\\
1.5&12&90&2.0&12&100\\
2.0&15&120&2.5&15&125\\
2.5&18&150&3.0&18&150\\
3.0&21&180&3.5&21&175\\
3.5&24&210&4.5&24&200\\
4.5&27&240&5.5&27&220\\
5.5&30&-&6.5&30&240\\
6.5&33&-&7.5&33&-\\
7.5&36&-&8.5&36&-\\
8.5&-&-&9.5&42&-\\
9.5&-&-&10.5&47&-\\
10.5&-&-&-&-&-\\
11.5&-&-&-&-&-\\
\hline
\hline
\end{tabular}
\caption{Grid points used in this work for sampling the
  reactant ($P(E_{\text{trans}}), P(v), P(j)$) and product state
  distributions ($ P(E_{\text{trans}}'), P(v'), P(j')$).}
\label{sifig:grid_table}
\end{table}%

\newpage
\section{Statistical Evaluation}
\label{si:stat_eval}
For statistical evaluation, NNs in the F- and G-based models were
trained on 10 independent random splits of $N_{\rm tot}$ into $N_{\rm
  train}$, $N_{\rm valid}$ and $N_{\rm test}$ for Set1 as well as Set1
and Set3, respectively. For each of the 10 resulting F- and G-based
models, ${\rm RMSD}_{\text{NN}}$, $R^2_{\text{NN}}$ and ${\rm
  RMSD}_{\text{QCT}}$, $R^2_{\text{QCT}}$ values were evaluated over
the test set and subsequently the corresponding mean and standard
deviation were calculated. These results are displayed in Table
\ref{sifig:statistical_eval}. Taking the reported standard deviations
as a reference, we can be confident that any performance difference
between two models larger than $\sim 0.0001$ is not solely of
statistical nature. In particular, we expect this to apply to
approaches and data sets other than the ones reported in Table
\ref{sifig:statistical_eval} and will therefore refer to this estimate
throughout this work.\\

\begin{table}[ht]
\centering
\begin{tabular}[t]{l||c|c|c}
\hline
\hline
DTD model&F-NN&G-NN&G-NN\\
\hline
Training \& test&Set1&Set1&Set3\\
${\rm RMSD}_{\rm NN}$&$0.00072\pm0.00005$&$0.00089\pm0.00002$&$0.00092\pm0.00003$\\
${R^2}_{\rm NN}$&$0.99948\pm0.00011$&$0.99930\pm0.00005$&$0.99905\pm0.00010$\\
${\rm RMSD}_{\rm QCT}$&$0.00142\pm0.00004$&$0.00107\pm0.00003$&$0.00106\pm0.00004$\\
${R^2}_{\rm QCT}$&$0.99816\pm0.00014$&$0.99901\pm0.00005$&$0.99882\pm0.00011$\\
\hline
\hline
\end{tabular}
\caption{Performance measures (${\rm RMSD}_{\rm NN}$, ${R^2}_{\rm
    NN}$, ${\rm RMSD}_{\rm QCT}$ and ${R^2}_{\rm QCT}$) for
  statistical evaluation of F- (F-NN) and G-based models (G-NN)
  trained and tested on Set 1/3.}
\label{sifig:statistical_eval}
\end{table}%

\section{Statistical Measures}
The RMSD$_{\rm QCT}$ and $R^2_{\rm QCT}$ values of all G-based models
and variants, applied to the different data sets are summarized in
Table \ref{sitab:perfomance_measures}.\\

\begin{table}[h!]
\centering
\begin{tabular}[t]{l|cccc}
\hline
\hline
DTD model&Training&Test&${\rm RMSD}_{\rm QCT}$&${R^2}_{\rm QCT}$\\
\hline
G-NN&Set1&Set1&0.0010&0.9991\\
\hline
G-NN&Set1&Set2&0.0016&0.9974\\
G-NN&Set2&Set2&0.0011&0.9990\\
G$'_{1}$-NN&Set2&Set2&0.0022&0.9878\\
G$'_{2}$-NN&Set2&Set2&0.0011&0.9990\\
G($\bm{T}$)-NN&Set2&Set2&0.0011&0.9990\\
G($\bm{\mu}$)-NN&Set2&Set2&0.0011&0.9990\\
\hline
G-NN&Set3&Set3&0.0011&0.9989\\
G$'_{2}$-NN&Set3&Set3&0.0013&0.9984\\
G$'_{3}$-NN&Set3&Set3&0.0011&0.9988\\
G($\bm{w},\bm{T}$)-NN&Set3&Set3&0.0011&0.9988\\
G($\bm{\mu}$)-NN&Set3&Set3&0.0020&0.9959\\
G($\bm{\mu},\bm{\sigma}$)-NN&Set3&Set3&0.0011&0.9987\\
\hline
G-NN&Set3&Set2&0.0012&0.9988\\
G$'_{3}$-NN&Set3&Set2&0.0012&0.9987\\
G($\bm{w},\bm{T}$)-NN&Set3&Set2&0.0125&0.8907\\
G($\bm{\mu},\bm{\sigma}$)-NN&Set3&Set2&0.0012&0.9988\\
\hline
G-NN&Set3&Set3A&0.0009&0.9991\\
G$'_{3}$-NN&Set3&Set3A&0.0009&0.9990\\
G($\bm{\mu},\bm{\sigma}$)-NN&Set3&Set3A&0.0011&0.9988\\
\hline
\hline
\end{tabular}
\caption{Performance measures (${\rm RMSD}_{\rm QCT}$ and ${R^2}_{\rm
    QCT}$) of all G-based models and variants considered in this
  work. Here, G$'_{i}$-NN denoted the G$'$-based model with $i$ grid
  points per reactant state distribution. The column labelled training
  denotes the data set on which the model was trained on, whereas the
  test column specifies the data set whose test set was used to
  calculate these performance measures. The number of significant
  digits being reported is based on the findings of Section
  \ref{si:stat_eval} in the SI.}
\label{sitab:perfomance_measures}
\end{table}%

\newpage
\section{G$'$-based models for Equilibrium Reactant State Distributions}
\label{si:limits_reduced_G_based_models}
G$'$-based models are variants of G-based models which use a
significantly reduced number of grid points per reactant state
distribution. Consider the case of a G$'$-based model, developed for
equilibrium reactant state distributions such as Set2.\\

\noindent
As these are equilibrium distributions they are characterized by a
corresponding temperature once the system-specific parameters are
fixed (here these are the energies of the rovibrational state of the
diatom). Thus, providing the value (``amplitude'') of the distribution
at a single, "suitably chosen" grid point uniquely determines the
corresponding distribution in the absence of noise and it is expected
that this suffices as input to accurately predict the corresponding
product state distributions. This is supported by the finding that
G($\bm{\mu}$)- and G($\bm{T}$)-based models yield accurate predictions
once ($\bm{\mu}$ and $\bm{T}$) are specified. However, it is important
that the grid point chosen does not correspond to a crossing point
between two equilibrium distributions at two different temperatures.\\

\noindent
In practice, however, the reactant state distributions considered in
this work suffer from noise due to finite sample statistics. In the
presence of noise the grid points should be placed at locations where
the difference between equilibrium distributions at different
temperatures is largest. Grid points located where this difference is
small, such as at the tail of these distributions, further raises the
difficulty of distinguishing between equilibrium distributions at
different temperatures when taking noise into account. Consequently,
the presence of noise serves as an explanation on why G$'$-based
models developed for equilibrium reactant state distributions of Set2
suffer from a significant drop in the prediction accuracy when the
number of grid points is reduced to a single point per distribution.\\

\newpage
\section{Additional Figures}
\begin{figure}[h]
\begin{center}
\includegraphics[width=0.99\textwidth]{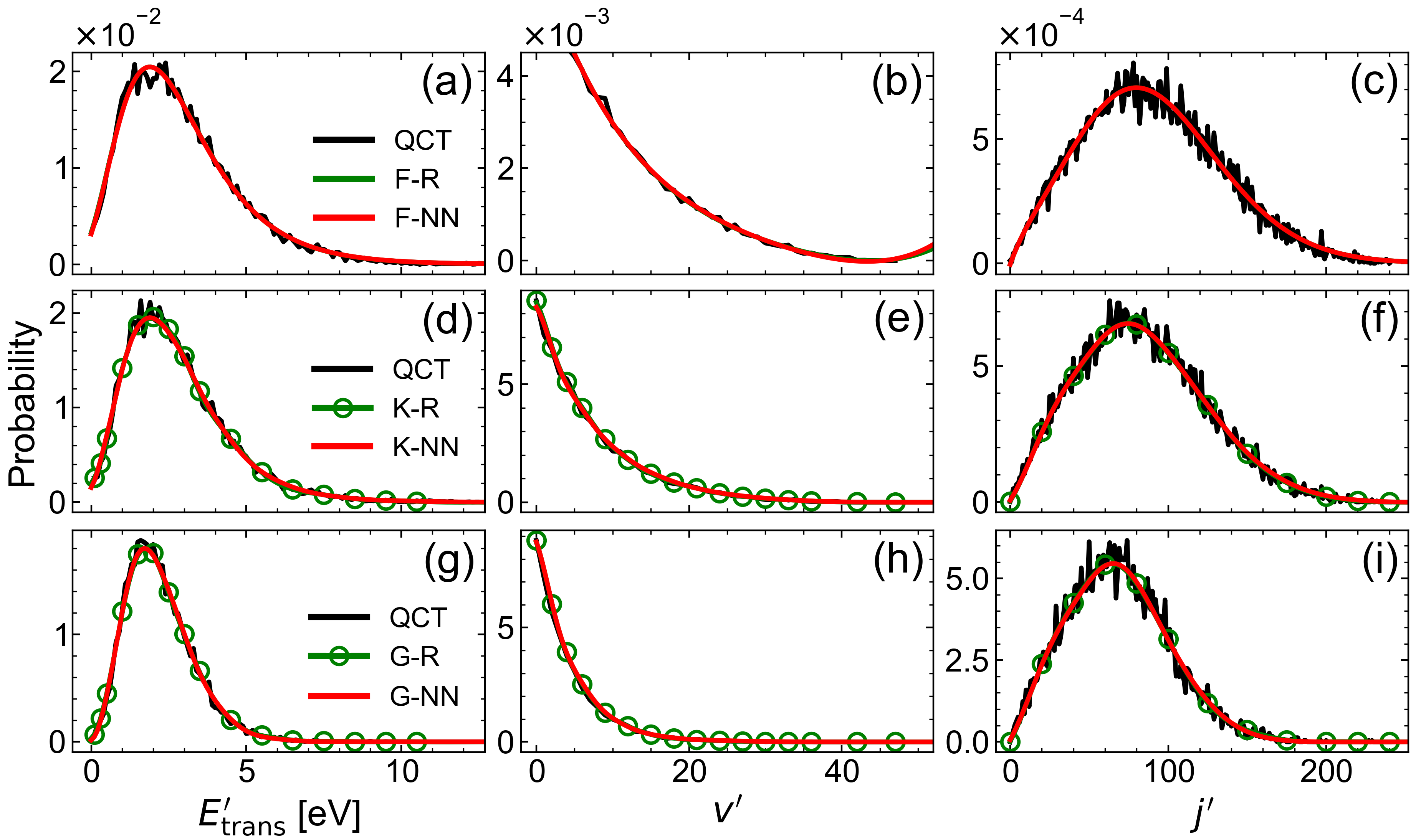}
\caption{Product state distributions obtained by explicit QCT
  simulations (QCT) as well as the corresponding references (-R) and
  predictions (-NN) obtained in the (a-c) F-based (F-R, F-NN), (d-f)
  K-based (K-R, K-NN) and (g-i) G-based approaches (G-R,
  G-NN). Furthermore, the amplitudes to construct the reference
  RKHS-based representations in the K- and G-based approaches are
  displayed (circles). The data sets considered here are from the test
  set of Set1 and result in predictions with the largest $R^2_{\rm
    NN}$ value in the test set: (a-c) $\bm{T}=($17750 K, 12500 K,
  12500 K), RMSD$_{\text{NN}}=0.0002$, $R^2_{\text{NN}}=0.99997$,
  (d-f) $\bm{T}=($16750 K, 8000 K, 8000 K), RMSD$_{\text{NN}}=0.0006$,
  $R^2_{\text{NN}}=0.9997$, (g-i) $\bm{T}=($9750 K, 5000 K, 5000 K),
  RMSD$_{\text{NN}}=0.0005$, $R^2_{\text{NN}}=0.9998$.}
\label{sifig:NN_comparison_accurate}
\end{center}
\end{figure}
\newpage
\begin{figure}[h]
\begin{center}
\includegraphics[width=0.99\textwidth]{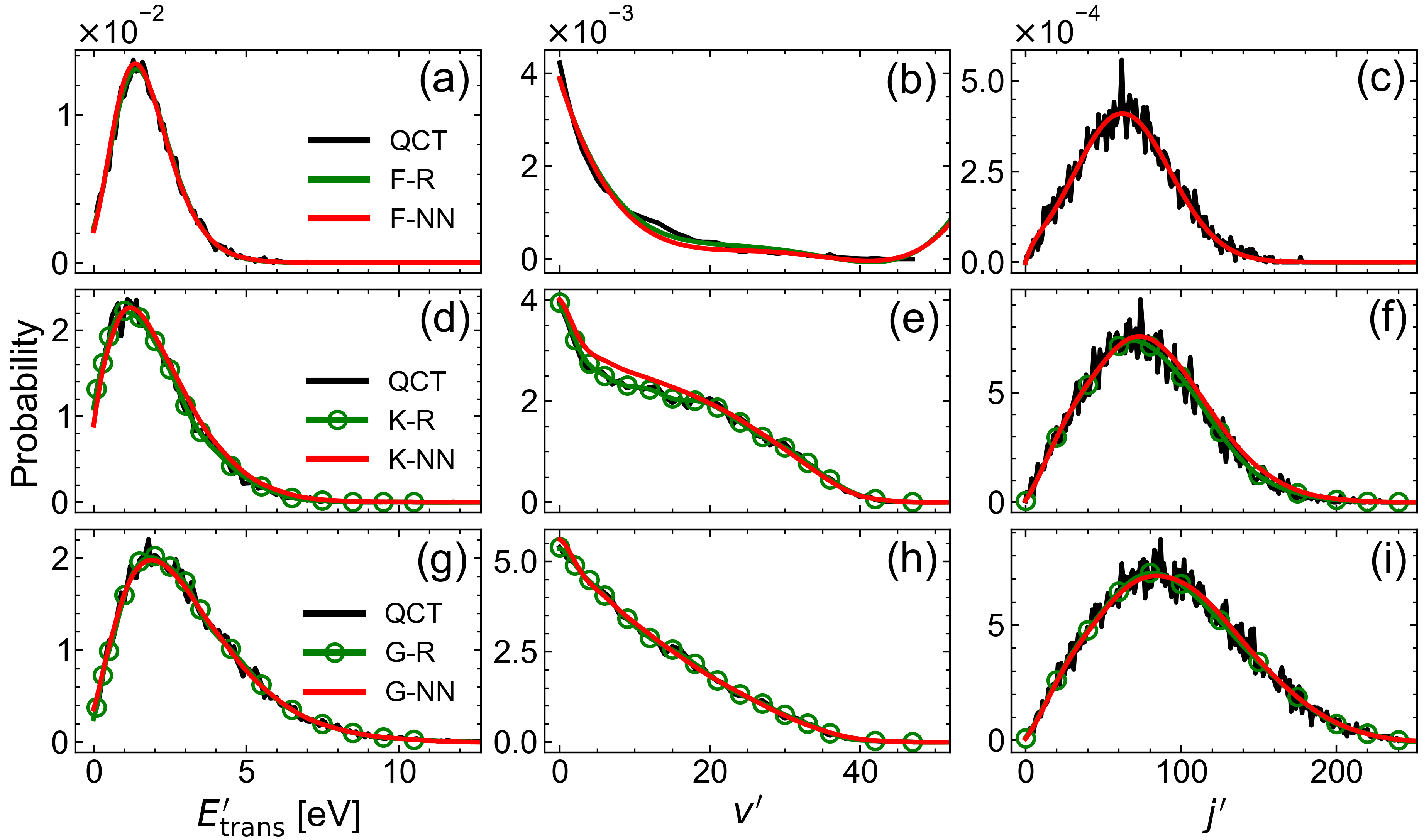}
\caption{Product state distributions obtained by explicit QCT
  simulations (QCT) as well as the corresponding references (-R) and
  predictions (-NN) obtained in the (a-c) F-based (F-R, F-NN), (d-f)
  K-based (K-R, K-NN) and (g-i) G-based approaches (G-R,
  G-NN). Furthermore, the amplitudes to construct the reference
  RKHS-based representations in the K- and G-based approaches are
  displayed (circles). The data sets considered here are from the test
  set of Set1 and result in predictions with the smallest $R^2_{\rm
    NN}$ value in the test set: (a-c) $\bm{T}=($5000 K, 8000 K, 8000
  K), RMSD$_{\text{NN}}=0.0020$, $R^2_{\text{NN}}=0.9973$, (d-f)
  $\bm{T}=($6750 K, 18750 K, 18750 K), RMSD$_{\text{NN}}=0.0036$,
  $R^2_{\text{NN}}=0.9887$, (g-i) $\bm{T}=($19000 K, 18750 K, 18750
  K), RMSD$_{\text{NN}}=0.0015$, $R^2_{\text{NN}}=0.9984$.}
\label{sifig:NN_comparison_inaccurate}
\end{center}
\end{figure}

\newpage

\begin{figure}[h]
\includegraphics[width=0.65\textwidth]{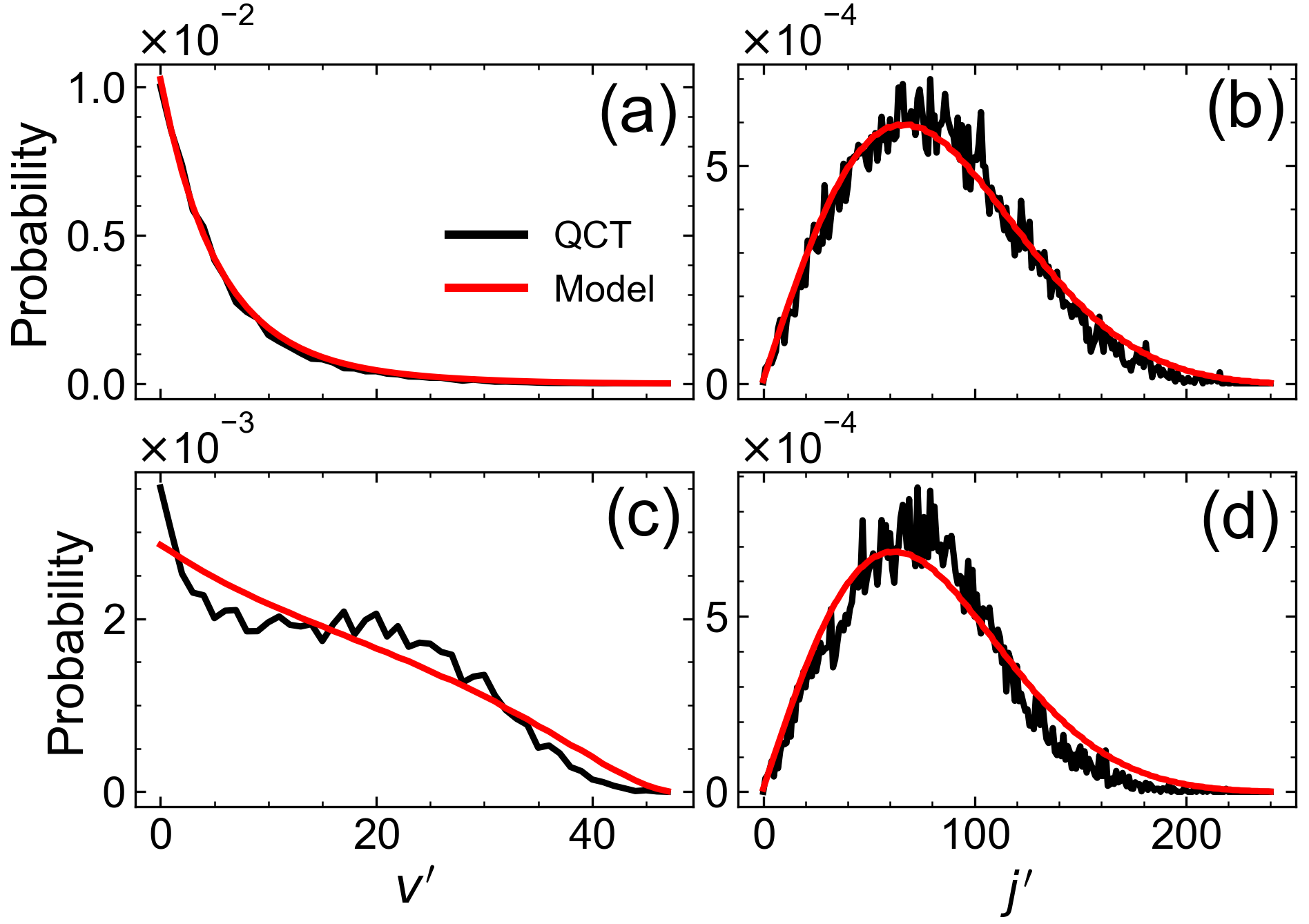}
\caption{$P(v')$ and $P(j')$ obtained by explicit QCT simulations
  (QCT) as well as the corresponding fits to the parametric surprisal
  model\cite{schwartz:2018} (Model). The data sets considered here are
  from Set1: (a-b) $\bm{T}$ = (20000 K, 5000 K, 5000 K), (c-d)
  $\bm{T}$ = (5500 K, 20000 K, 20000 K). While the model closely
  matches the QCT data for (a-b), it is insufficiently described by
  the model for (c-d), in particular $P(v')$.}
\label{sifig:Schwarzentruber}
\end{figure}

\newpage

\begin{figure}[h!]
\begin{center}
\includegraphics[width=0.99\textwidth]{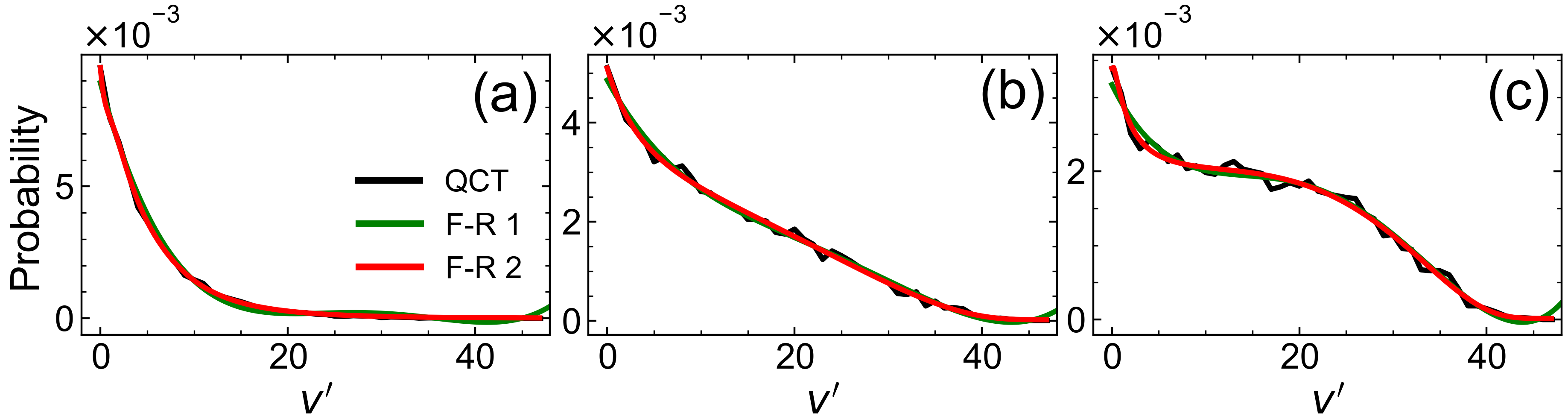}
\caption{$P(v')$ obtained by explicit QCT simulations (QCT) as well as
  the corresponding references obtained in the F-based approach using
  Eq.~6 (F-R 1) and Eq.~12
  (F-R 2). The data sets considered here are from Set1: (a) $\bm{T}$ =
  (12500 K, 5750 K, 5750 K), (b) $\bm{T}$ = (9500 K, 16000 K, 16000
  K), (c) $\bm{T}$ = (5750 K, 19250 K, 19250 K). These results
  illustrate that Eq.~12 leads to a better
  quality of fit compared to Eq.~6.}
\label{sifig:alternative_function_based_model}
\end{center}
\end{figure}
\begin{figure}[h!]
\begin{center}
\includegraphics[width=0.99\textwidth]{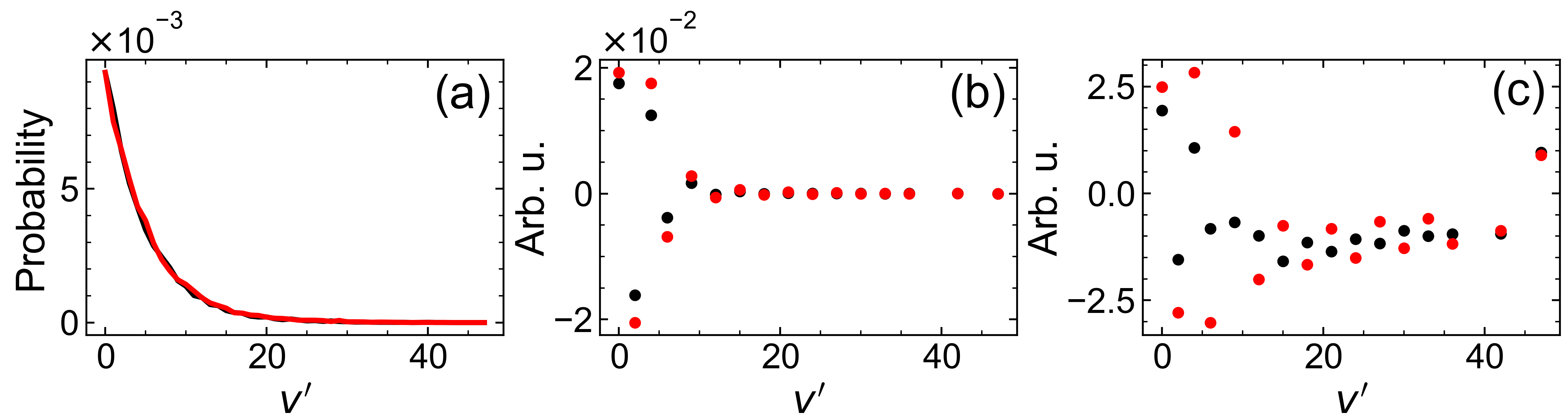}
\caption{(a) $P(v')$ obtained by explicit QCT simulations for
  ($T_{\text{trans}}=12000$ K, $T_{\text{rovib}}=5250$ K; black) and
  ($T_{\text{trans}}=12000$ K, $T_{\text{rovib}}=5500$ K; red) with
  similar shape. (b-c) Corresponding featurizations using the K-based
  approach. Here, the displayed features (kernel coefficients) are (b)
  non-standardized and (c) standardized. These results illustrate that
  positive and negative kernel coefficients can cancel. Hence,
  different combinations of kernel coefficients are able to model
  similarly shaped distributions, here $P(v')$.}
\label{sifig:pv_kernelcoeff}
\end{center}
\end{figure}

\newpage

\begin{figure}[h]
\begin{center}
\includegraphics[width=0.99\textwidth]{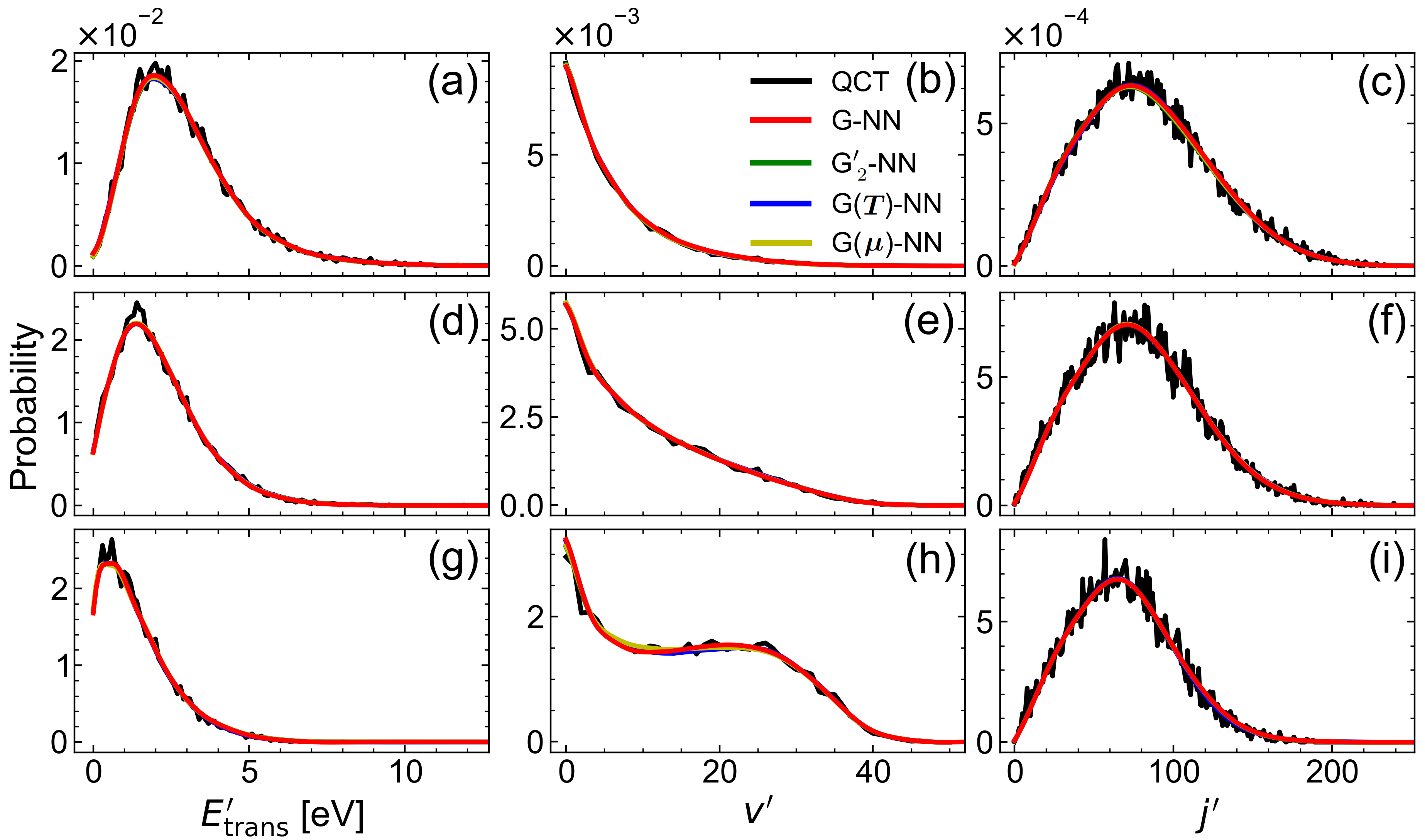}
\caption{Product state distributions obtained from explicit QCT
  simulations (QCT), as well as the corresponding predictions of the
  G- (G-NN), G$'_{2}$- (G$'_{2}$-NN), G($\bm{T}$)- (G($\bm{T}$)-NN)
  and G($\bm{\mu}$)-based models (G($\bm{\mu}$)-NN). G$'_{2}$-NN uses
  2 grid points per reactant state distribution (see main text). The
  data sets considered here are from the test set of Set2: (a-c)
  $\bm{T}$ = (20000 K, 7000 K, 5000 K), (d-f) $\bm{T}$ = (10000 K,
  14000 K, 9000 K), (g-i) $\bm{T}$ = (5000 K, 20000 K, 8000 K).}
\label{sifig:set2_differentinputs}
\end{center}
\end{figure}

\newpage

\begin{figure}[h]
\begin{center}
\includegraphics[width=0.6\textwidth]{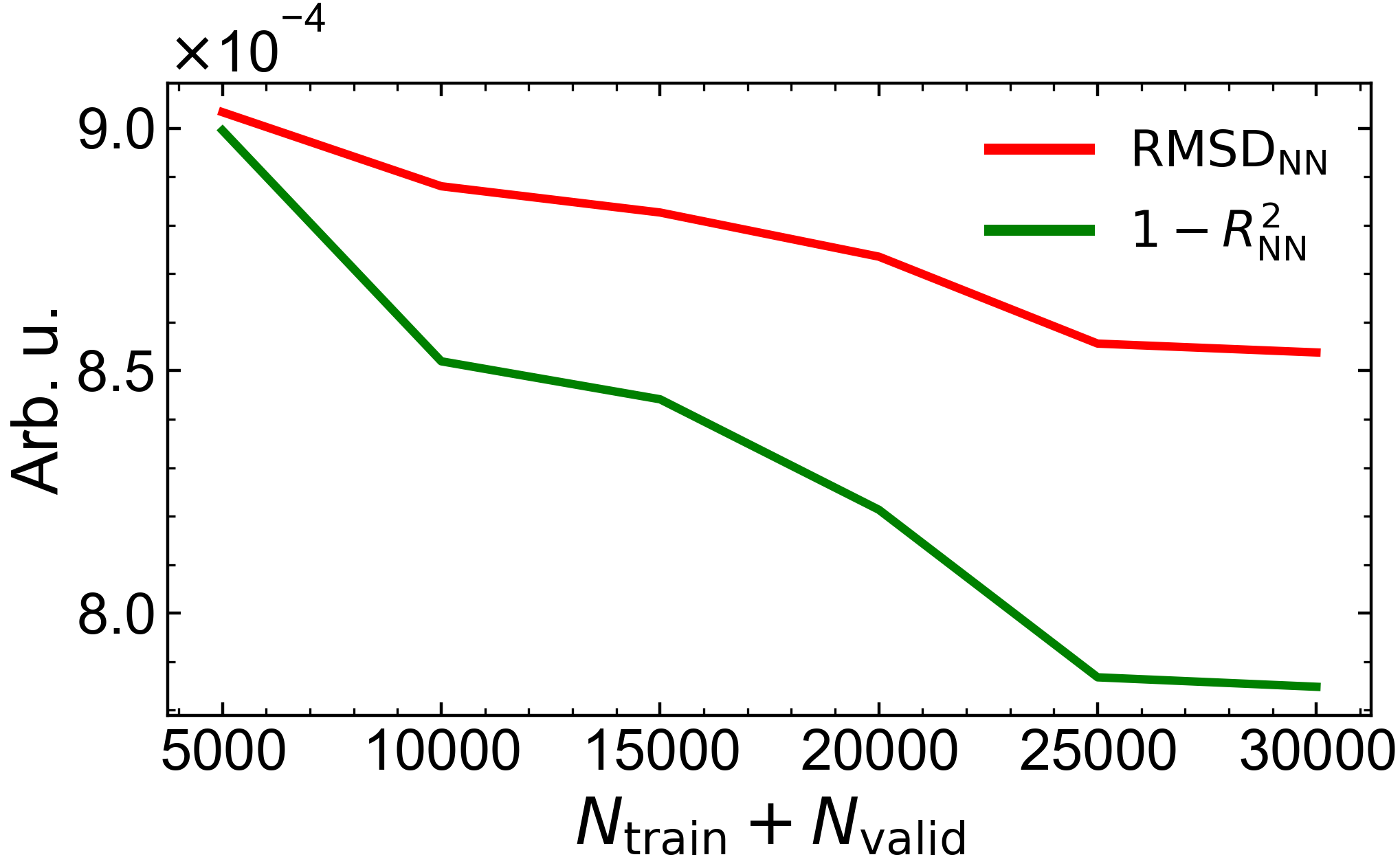}
\caption{Learning curve for the G-based model trained and evaluated on
  Set3 with a variable size of the training and validation sets. Here,
  both RMSD$_{\rm NN}$ and $1-R^2_{\rm NN}$ are measures for the NN
  prediction accuracy. The NN prediction accuracy did not
  significantly increase when $N_{\text{train}}+N_{\text{valid}}$ was
  increased from 5000 to 30000 in increments of 5000.}
\label{sifig:probmodel_learningcurve}
\end{center}
\end{figure}
\begin{table}[h]
\centering
\begin{tabular}[t]{l|cc}
\hline
\hline
Panels&$w_{n}/w_{\rm tot}$&$\bm{T}$ [K]\\
\hline
(a-c)&2/5&$(11750, 6250, 6250)$\\
&2/5&$(7750,8250,8250)$\\
&1/5&$(16750, 19000, 19000)$\\
\hline
(d-f)&1/2&$(10000, 14750, 14750)$\\
&1/2&$(20000, 18000, 13000)$\\
\hline
(g-i)&2/3&$(19500, 17750,17750)$\\
&1/3&$(5000, 7000, 19000)$\\
\hline
\hline
\end{tabular}
\caption{Normalized weights $w_{n}/w_{\rm tot}$ and sets of
  temperatures $\bm{T}=(T_{\rm trans}, T_{\rm vib}, T_{\rm rot})$
  characterizing the data sets displayed in Figure~8.}
\label{sifig:probmodel_differentinputs_R2_table}
\end{table}%

\newpage

\begin{table}[h]
\centering
\begin{tabular}[t]{l|cc}
\hline
\hline
Panels&$w_{n}/w_{\rm tot}$&$\bm{T}$ [K]\\
\hline
(a-c)&39/77&$(15000,19750, 19750)$\\
&6/77&$(14250, 9500 ,9500)$\\
&32/77&$(10250, 10250, 10250)$\\
\hline
(d-f)&71/611&$(15000, 19750 ,19750)$\\
&82/611&$(15250 ,12250 ,12250)$\\
&84/611&$(14250 ,18500 ,18500)$\\
&16/611&$(9500,19750 ,19750)$\\
&24/611&$(10000, 14000 ,15000)$\\
&23/611&$(20000, 11000 ,16000)$\\
&95/611&$(9750 ,20000,20000)$\\
&89/611&$(19500 ,17750 ,17750)$\\
&51/611&$(14500 ,19500,19500)$\\
&76/611&$(10250, 10250,10250)$\\
\hline
(g-i)&18/77&$(17500, 7500, 7500)$\\ 
&32/77&$(7750 ,16000,16000)$\\
&5/77&$(10000, 18000, 8000)$\\
&22/77&$(18750 ,5000 ,5000)$\\
\hline
\hline
\end{tabular}
\caption{Normalized weights $w_{n}/w_{\rm tot}$ and sets of
  temperatures $\bm{T}=(T_{\rm trans}, T_{\rm vib}, T_{\rm rot})$
  characterizing the data sets displayed in Figure~9.}
\label{sifig:probmodel_diverse_differentinputs_R2_table}
\end{table}%

\clearpage
\bibliography{refs}